    \documentclass[8pt, final, twocolumn]{extarticle}

    \usepackage[]{geometry} 
    \usepackage[utf8]{inputenc}
    \usepackage[T1]{fontenc}

    \usepackage[english]{babel}

    \usepackage{xcolor}
    
    \usepackage{amsmath}
    \usepackage{amssymb}

    \usepackage{bm}
 
    \usepackage[binary-units=true]{siunitx} 

    \allowdisplaybreaks   

	\usepackage{authblk}	
    
    \usepackage[font={footnotesize},labelsep=period]{caption}   
    \usepackage{subcaption}
    
    \usepackage{pgfplots}\pgfplotsset{compat=1.14}
    \usepackage{Util/UTIL_PGFplotTEMF/pgfplots-private}
    \usepackage{pgfplotstable}
        
    \usepgfplotslibrary{units} 
    \usepgfplotslibrary{fillbetween}
    \usetikzlibrary{shapes,arrows,positioning,decorations.markings}

\newcommand{\ApplyCommonBarPlotSettings}[0]{
    ybar = 1pt,
    temfbarplot,
    bar width = 1pt,
}

\newcommand{\FigWorkingPrinciple}[0]{
    \begin{tikzpicture}
    
        \node
            [inner sep=0pt] 
            at (0,0)
            {\includegraphics[width=8.5cm]{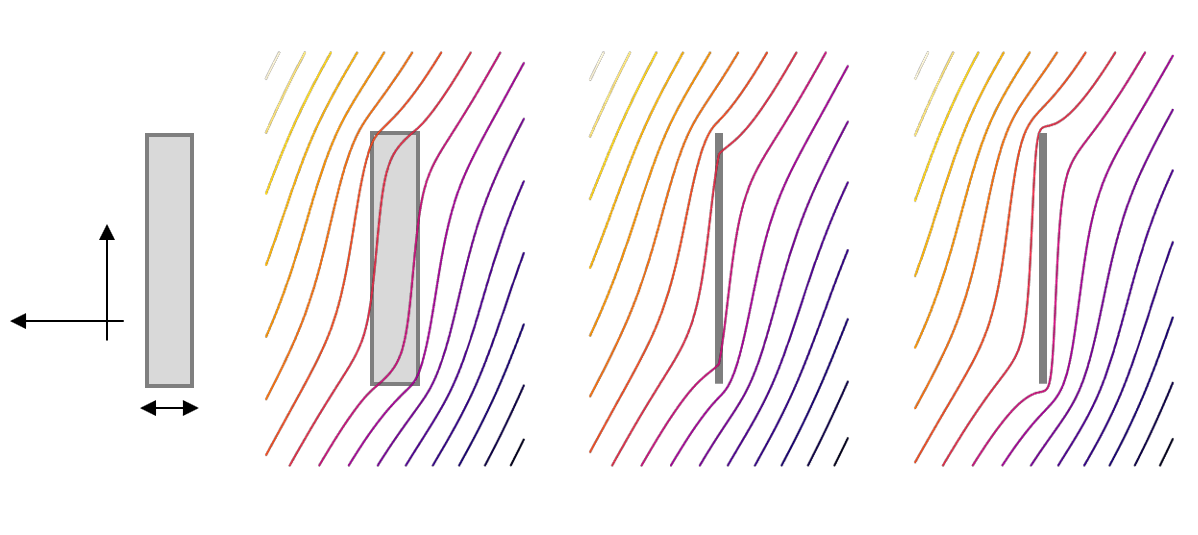}};       
        \node 
            [anchor=center]
            at (-3.5, -1.7) 
            {\scriptsize{a)}}; 
        \node 
            [anchor=center]
            at (-1.45, -1.7) 
            {\scriptsize{b)}}; 
        \node 
            [anchor=center]
            at (0.8, -1.7) 
            {\scriptsize{c)}}; 
        \node 
            [anchor=center]
            at (3.1, -1.7) 
            {\scriptsize{d)}};
        
        \node 
            [anchor=center]
            at (-1.45, 1.7) 
            {\scriptsize{$\SCAl{\ITdelta}{\RM{s}},\SCA{\ITrho}\gg0$}}; 
        \node 
            [anchor=center]
            at (0.8, 1.7) 
            {\scriptsize{$\SCA{\ITrho}\gg0,\ \SCAl{\ITdelta}{\RM{s}}\to0$}}; 
        \node 
            [anchor=center]
            at (3.1, 1.7) 
            {\scriptsize{$\SCAl{\ITdelta}{\RM{s}},\SCA{\ITrho}\to0$}};
        
        \node 
            [anchor=center]
            at (-3.05, -1.2) 
            {\scriptsize{$\SCAl{\ITdelta}{\RM{s}}$}}; 
        \node 
            [anchor=center]
            at (-3.05, -0.5) 
            {\scriptsize{$\SCA{\ITrho}$}}; 
        \node 
            [anchor=center]
            at (-4.0, -0.2) 
            {\scriptsize{$\PERP$}}; 
        \node 
            [anchor=center]
            at (-3.7, 0.2) 
            {\scriptsize{$\PARA$}};              
	\end{tikzpicture}
}

\newcommand{\FigLineMagnetCurrentProfile}[1]{
    \begin{tikzpicture}
        \pgfplotstableread [col sep=comma]{#1}\datatable 
    	    \begin{axis}[    		
            temflineplot,
            legend pos  = north east,
            title       = Current profile for the dipole magnet MCB24,
            width       = 8.5cm,
            height      = 6cm,
            xmin        = 0, 
            xmax        = 300,
            ymin        = -100, 
            ymax        = 400,
            xticklabels = {,,},
            ytick       = {-100,0,...,400},
            xlabel      = Time, 
            ylabel      = $\RM{Magnet\ current}$,
            x unit      = \si{{a.u.}}, 
            y unit      = \si{\ampere},           
            ]   
        
        \addplot+ [mark=none] table[x=xt_mag1,y=yt_mag1]{\datatable};
        \pgfplotsset{cycle list shift=-1};
        \addplot+ [mark=none] [no marks,dotted] table[x=xt_mag2,y=yt_mag2]{\datatable};
        \pgfplotsset{cycle list shift=-2};
        \addplot+ [mark=none] table[x=xt_mag3,y=yt_mag3]{\datatable};
        \pgfplotsset{cycle list shift=-3};
        \addplot+ [no marks,dotted] table[x=xt_mag4,y=yt_mag4]{\datatable};
        \pgfplotsset{cycle list shift=-4};
        \addplot+ [mark=none] table[x=xt_mag5,y=yt_mag5]{\datatable};
        \pgfplotsset{cycle list shift=-5};
        \addplot+ [no marks,dotted] table[x=xt_mag6,y=yt_mag6]{\datatable};  
        
        \addplot [mark=none, dashed, gray!100]
            coordinates {(120,-200) (120,500)};  
        \addplot [mark=none, dashed, gray!100]
            coordinates {(200,-200) (200,500)};  
        \addplot [mark=none, dashed, gray!100]
            coordinates {(280,-200) (280,500)};  
       
        \node at (60, -50) [align=center]
            {\scriptsize{$\RM{Pre}$-$\RM{cycle}$}}; 
        \node at (160, -50) [align=center]
            {\scriptsize{$\RM{Test}_{{i}}$}}; 
        \node at (240, -50) [align=center] 
            {\scriptsize{$\RM{Test}_{{i}+1}$}}; 
         \node at (170.5,100) [align=center]
            {\scriptsize{$\RM{Data}_{{i}}$}}; 
        \node at (250.5,100) [align=center] 
            {\scriptsize{$\RM{Data}_{{i}+1}$}};          
            
        \draw [->,thick]
            (180.5,80) -- (180.5,40);
        \draw [->,thick]
            (260.5,80) -- (260.5,40);   

    \end{axis}   
    \end{tikzpicture}
}

\newcommand{\figHALOVisualInspection}[0]{
    \begin{tikzpicture}
    
        \node
            [inner sep=0pt] 
            at (0,0)
            {\includegraphics[width=8.5cm]{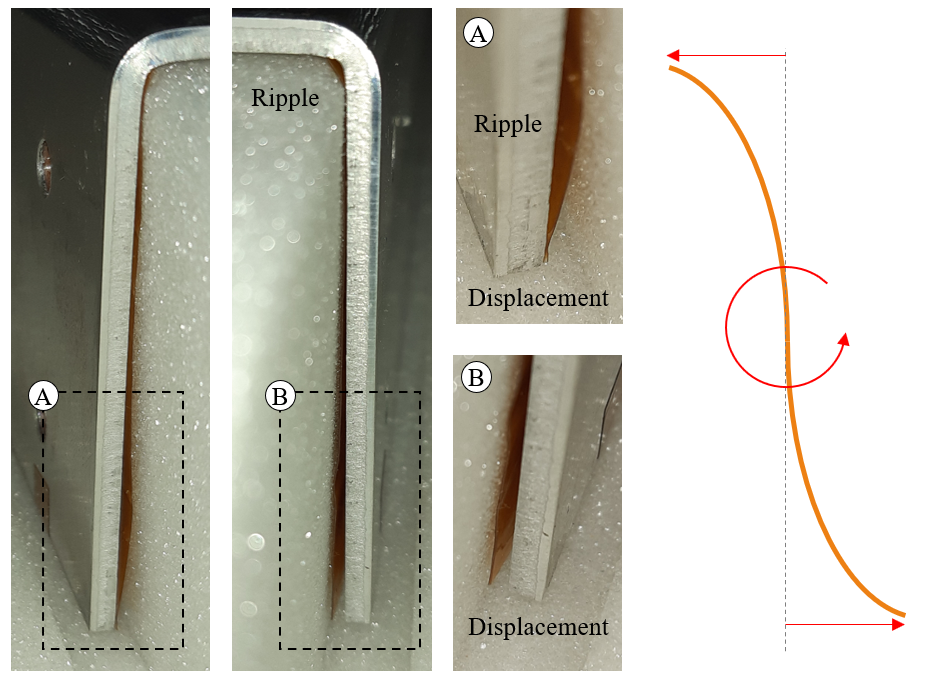}};
        \node 
            [anchor=west]
            at (1.9, 2.8) 
            {\scriptsize{$\pm\SCAl{\ITvarepsilon}{\RM{up}}$}};
        \node 
            [anchor=west]
            at (2.9, -2.8) 
            {\scriptsize{$\pm\SCAl{\ITvarepsilon}{\RM{dn}}$}};  
        \node 
            [anchor=west]
            at (3.2, 0.3) 
            {\scriptsize{$\pm\SCAl{\ITvarepsilon}{\RMtheta}$}};    
	\end{tikzpicture}
}

\newcommand{\FigBHcurvePureIron}[1]{
    	\begin{tikzpicture}
    	\pgfplotstableread [col sep=comma]{#1}\datatable 
		\begin{semilogxaxis}[	
            temflineplot,
            legend pos  = north east,
            title       = Constitutive relation for iron,
            width       = 8.5cm,
            height      = 6cm,
            xmin        = 1e1, 
            xmax        = 1e5,
            ymin        = 0, 
            ymax        = 2.5,
            ytick       = {0,0.5,...,2.5},
            xlabel      = Magnetic field, 
            ylabel      = Magnetic flux density,
            x unit      = A/m, 
            y unit      = \SI{}{\tesla},       
            ]                
        \addplot+ table[x={xH_iron},y={yB_iron}]{\datatable};
        \label{plot_1_y1}   
	    \end{semilogxaxis}		
		\begin{semilogxaxis}[	
            temflineplot,
		    axis y line*= right,
		    xmajorgrids = false,
	    	ymajorgrids = false,
            legend pos  = north east,
            width       = 8cm, 
            height      = 6cm,
            xmin        = 1e1, 
            xmax        = 1e5,
            ymin        = 0, 
            ymax        = 5000,
            ytick      = {0,1e3,...,5e3},
            xlabel      = , 
            ylabel      = $\SCAl{\ITmu}{\RM{r}}$,
            x unit      = , 
            y unit      = \si{-}, 
            hide x axis
            ] 
	        \addlegendimage{/pgfplots/refstyle=plot_1_y1}
	        \addlegendentry{$\RM{B}$}
	        
            \pgfplotsset{cycle list shift=+1}
            \addplot+ table[x={xH_iron},y={ymur_iron}]{\datatable};
            \label{plot_1_y2}
		
            \addlegendimage{/pgfplots/refstyle=plot_1_y2}
            \addlegendentry{$\SCAl{\ITmu}{\RM{r}}$}	 
            
		\end{semilogxaxis}				
	\end{tikzpicture}
}

\newcommand{\FigLiftingFunction}[1]{
    \begin{tikzpicture}
        \pgfplotstableread [col sep=comma]{#1}\datatable 
    	    \begin{axis}[    		
            temflineplot,
            legend pos  = south east,
            title       = Lifting function $\SCAl{f}{\RM{l}}$ at $\SI{77}{\kelvin}$,
            width       = 8.5cm,
            height      = 6cm,
            xmin        = 0, 
            xmax        = 180,
            ymin        = 0, 
            ymax        = 1.2,
            xtick       = {0,30,...,180},,            
            ytick       = {0,0.2,...,1.4},
            xlabel      = Field angle $\ITtheta$, 
            ylabel      = ,
            x unit      = \si{\degree}, 
            y unit      = \si{-},           
            ]   
        
        \addplot+ table[x={x_theta},y={y_ic_self}]{\datatable};
        \addplot+ table[x={x_theta},y={y_ic_100mT}]{\datatable};

        \node 
		    [above right] at (rel axis cs:0.05,0.05) 
			{\fcolorbox{gray!100}{white}{\includegraphics[width=2.5cm]{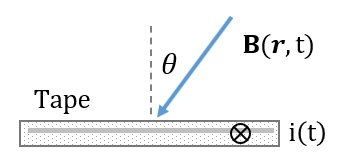}}}; 

        \legend{self field,   
                $\SI{100}{\milli\tesla}$,
                }        

    \end{axis}   
    \end{tikzpicture}
}

\newcommand{\FigLineHALOdeformation}[1]{
    \begin{tikzpicture}
        \pgfplotstableread [col sep=comma]{#1}\datatable 		
    		\begin{axis}[    		
            temflineplot,
            legend style={at={(0.26,0.85)},anchor=west},
            title       = Geometrical error of HTS screens,
            width       = 8.5cm,
            height      = 6cm,
            xmin        = -60, 
            xmax        = 60,
            ymin        = -40, 
            ymax        = 40,
            xtick       = {-80,-60,...,+80},
            ytick       = {-60,-40,...,+60},
            xlabel      = $\SCA{x}$, 
            ylabel      = $\SCA{y}$,
            x unit      = \si{\mm}, 
            y unit      = \si{\mm},
            ]     
        
        \addplot+ [mark repeat=39] table[x=xl_v01,y=yl_v01]{\datatable}; 
        \pgfplotsset{cycle list shift=-1};
        \addplot+ [mark repeat=39] table[x=xr_v01,y=yr_v01,forget plot]{\datatable}; 
        \pgfplotsset{cycle list shift=+0};
        \addplot+ [mark repeat=39] table[x=xl_v02,y=yl_v02]{\datatable}; 
        \pgfplotsset{cycle list shift=-1};
        \addplot+ [mark repeat=39] table[x=xr_v02,y=yr_v02,forget plot]{\datatable}; 
        \addplot [mark=none, black] table[x=xl_v00,y=yl_v00]{\datatable};  
        \addplot [mark=none, black,forget plot] table[x=xr_v00,y=yr_v00]{\datatable}; 
        \addplot   
            [mark=none, dashed, black] 
            table[x=xprobe,y=yprobe,col sep=comma] 
            {\datatable}; 
        
        \node 
            at (-43, -35) 
            {\scriptsize{Left screen}}; 
        \node 
            at (43, -35) 
            {\scriptsize{Right screen}}; 
        \node 
            at (-43, 35) 
            {\scriptsize{Error x10}}; 
        \node 
            at (43, 35) 
            {\scriptsize{Error x10}};                         
        \node 
            at (0, -20) 
            {\scriptsize{Rotating coil}};
        
        \addplot
            [mark=none, dotted, thin, gray!100]
            coordinates {(-100,0) (100,0)};    
            
        \legend{$1^{\RM{st}}$ HTS holder,   
			    $2^{\RM{nd}}$ HTS holder, 
			    Reference geometry}                  

    \end{axis}   
    \end{tikzpicture}
}

\newcommand{\FigBarMultipolesHALO}[5]{    
    \begin{tikzpicture}    
        \pgfplotstableread [col sep=comma]{#1}\datatable 
    		\begin{axis}[  
            \ApplyCommonBarPlotSettings{},
            xtick = data,
            xticklabels = {$\RM{b}_2$,$\RM{a}_2$,$\RM{b}_3$,$\RM{a}_3$,
                          $\RM{b}_4$,$\RM{a}_4$,$\RM{b}_5$,$\RM{a}_5$,
                          $\RM{b}_6$,$\RM{a}_6$},
            typeset ticklabels with strut,
            title       = {#2},
            width       = 8.5cm,
            height      = 6cm,
            ymin        = {#3}, 
            ymax        = {#4},	
            ytick       = {#3,{#3+#5},...,#4},
            ylabel      = $\RM{units}$,
            y unit      = \si{-},   
            ]     
        
        \addplot+ table[x=xMP_pos,y=yMP_300K_meas]{\datatable}; 
        \addplot+ table[x=xMP_pos,y=yMP_300K_sim]{\datatable};  
        \addplot+ table[x=xMP_pos,y=yMP_77K_meas]{\datatable}; 
        \addplot+ table[x=xMP_pos,y=yMP_77K_simPSO]{\datatable};
        \addplot+ table[x=xMP_pos,y=yMP_77K_sim]{\datatable}; 
           
        \legend{300K meas.,
			    300K sim.,
			    77K meas.,
			    77K sim. (curved),
			    77K sim. (flat),}	 			 
    \end{axis}    
    \end{tikzpicture}
}

\newcommand{\FigBarMultipolesScenarios}[6]{
    \begin{tikzpicture}
        \pgfplotstableread [col sep=comma]{#1}\datatable 
    	    \begin{axis}[    		
            \ApplyCommonBarPlotSettings,
            legend pos  = south east,
            xtick = data,
            xticklabels = {$\RM{b}_2$,$\RM{a}_2$,$\RM{b}_3$,$\RM{a}_3$,
                          $\RM{b}_4$,$\RM{a}_4$,$\RM{b}_5$,$\RM{a}_5$,
                          $\RM{b}_6$,$\RM{a}_6$},
            typeset ticklabels with strut,
            axis x line*=bottom,
            title       = {#4},
            width       = 8.5cm,
            height      = 6cm,
            ymin        = #2, 
            ymax        = #3,	
            ytick       = {#2,#2+5,...,#3},
            ylabel      = $\RM{units}$,
            y unit      = \si{-},   
            ]     
        \addplot+ table[x=xMP_pos,y=yMP_iron_meas]{\datatable}; 
        \addplot+ table[x=xMP_pos,y=yMP_iron_sim]{\datatable}; 
        \addplot+ table[x=xMP_pos,y=yMP_both_meas]{\datatable};
        \addplot+ table[x=xMP_pos,y=yMP_both_simPSO]{\datatable}; 
        \addplot+ table[x=xMP_pos,y=yMP_both_sim]{\datatable};

		\legend{300K meas.,
			    300K sim.,
			    77K meas.,
			    77K sim. (curved),
			    77K sim. (flat),}
			    			    
    \end{axis}    
    \end{tikzpicture}
}

\newcommand{\FigBarTHD}[4]{
    \begin{tikzpicture}
        \pgfplotstableread [col sep=comma]{#1}\datatable 
    	    \begin{axis}[  
            ybar = 1pt,
            temfbarplot,
            bar width = 4pt,
            legend pos  = north west,
            xtick={1,2,3,4,5},
            xticklabels = {medium,
                           low,
                           medium,
                           high,
                           check,},  
            xticklabel style = {align=center},
            axis x line*=bottom,
            title       = #4,
            width       = 8.5cm,
            height      = 6cm,
            ymin        = #2, 
            ymax        = #3,	
            ytick       = {#2,#2+10,...,#3},
            ylabel      = $\RM{units}$,
            y unit      = \si{-},   
            ]     
        \addplot+ table[x=xTHD_pos,y=yTHD_iron_meas]{\datatable}; 
        \addplot+ table[x=xTHD_pos,y=yTHD_iron_sim]{\datatable}; 
        \addplot+ table[x=xTHD_pos,y=yTHD_both_meas]{\datatable}; 
        \addplot+ table[x=xTHD_pos,y=yTHD_both_simPSO]{\datatable}; 
        \addplot+ table[x=xTHD_pos,y=yTHD_both_sim]{\datatable}; 	       
			    		
		\draw 
		    [mark=none, dashed, gray!100] 
		    (1.5,0) -- (1.5,50) ;		    
		 \node 
            at (1.30, 30) 
            [align=left,text width=1.5cm,rotate=90]
            {\scriptsize{$1^\RM{st}$ holder}}; 
		 \node 
            at (1.70, 30) 
            [align=left,text width=1.5cm,rotate=90]
            {\scriptsize{$2^\RM{nd}$ holder}};         
           
		\legend{300K meas.,
			    300K sim.,
			    77K meas.,
			    77K sim. (curved),
			    77K sim. (flat),}	 
        \end{axis}
    \end{tikzpicture}
}

\newcommand{\FigBarResultsExtrapolationGeomTenTesla}[1]{
    \begin{tikzpicture}
        \pgfplotstableread [col sep=comma]{#1}\datatable 
    	    \begin{axis}[    		
            ybar = 1pt,
            temfbarplot,
            bar width = 4pt,
            legend pos  = north east,
            xtick = data,
            xticklabels from table = {\datatable}{xTHD_label}, 
            axis x line*=bottom,
            title       = {{THD} factor},
            width       = 8.5cm,
            height      = 6cm,
            ymin        = 0, 
            ymax        = 12,	
            ytick       = {1,2,...,12},
            xlabel      = Number of tapes per layer, 
            ylabel      = $\RM{units}$,
            x unit      = \si{-}, 
            y unit      = \si{-},  
            ]     
        \addplot+ table[x={xTHD_idx},y={yTHD_2layers}]{\datatable}; 
        \addplot+ table[x={xTHD_idx},y={yTHD_4layers}]{\datatable}; 
        \addplot+ table[x={xTHD_idx},y={yTHD_8layers}]{\datatable}; 
        
		\draw 
		    [mark=none, dashed, red] 
		    (0,10) -- (3.5,10) ;		    
        \node 
            at (1.65, 10.5) 
            [align=left]
            {\scriptsize{Field error without screens}}; 
            
		\legend{4 layers,
			    8 layers,
			    16 layers,
			    }	
    \end{axis}   
    \end{tikzpicture}
}

\newcommand{\FigLineResultsExtrapolationTimeTenTesla}[1]{
    \begin{tikzpicture}
        \pgfplotstableread [col sep=comma]{#1}\datatable 
    	    \begin{loglogaxis}[    		
            temflineplot,
            legend pos  = north east,
            axis x line*=bottom,
            title       = {Drift of THD factor},
            width       = 8.5cm,
            height      = 6cm,
            xmin        = 1e-2, 
            xmax        = 1e+3,
            ymin        = 1e-4, 
            ymax        = 1e2,
            ytickten    = {-4,...,3},
            xlabel      = Time, 
            ylabel      = $\RM{units}$,
            x unit      = \si{\hour}, 
            y unit      = \si{-},           
            ]  
        
        \addplot+ table[x={x_t},y={y_5t_2l}]{\datatable};  
        \addplot+ table[x={x_t},y={y_6t_2l}]{\datatable}; 
        \addplot+ table[x={x_t},y={y_7t_2l}]{\datatable}; 
        \addplot+ table[x={x_t},y={y_8t_2l}]{\datatable}; 
        
        \pgfplotsset{cycle list shift=-4};   
        \addplot+ table[x={x_t},y={y_5t_8l}]{\datatable}; 
        \addplot+ table[x={x_t},y={y_6t_8l}]{\datatable}; 
        \addplot+ table[x={x_t},y={y_7t_8l}]{\datatable}; 
        \addplot+ table[x={x_t},y={y_8t_8l}]{\datatable}; 
        
		\draw 
		    [mark=none, dashed, red] 
		    (0.01,10) -- (10,10) ;		    
        \node 
            at (0.2, 20) 
            [align=left]
            {\scriptsize{Field error without screens}}; 
                
        \draw [decorate,decoration={brace,mirror,amplitude=5pt},xshift=0pt,yshift=0pt]
            (120,2e-3) -- (120,2e-2) 
            node [black,midway,xshift=0.7cm] {\scriptsize{16 layers}};    
        
        \draw [decorate,decoration={brace,mirror,amplitude=5pt},xshift=0pt,yshift=0pt]
            (120,3e-2) -- (120,3e-1) 
            node [black,midway,xshift=0.7cm] {\scriptsize{4 layers}};        
    	
    	\legend{5 tapes per layer,
			    6 tapes per layer,
			    7 tapes per layer,
			    8 tapes per layer,
			    }			    
    \end{loglogaxis}   
    \end{tikzpicture}
}

    \usepackage{booktabs}
    
    \usepackage{etoolbox}
    \AtBeginEnvironment{tabular}{\footnotesize}

    \usepackage{multirow}

\newcommand{\TabTapeSpecifications}[0]{
        \begin{tabular}{lccl}
            \toprule
            Parameter & Unit & Value & Description  \\
            \midrule
            Superpower Inc.                    &                       &     & Producer            \\
            SCS12050-AP                        &                       &     & Tape label          \\
            $\RMdelta_{\RM{w}}$                & [$\si{\milli\meter}$] & 12  & Tape width          \\
            $\RMdelta_{\RM{t}}$                & [$\si{\micro\meter}$] & 100 & Tape thickness      \\
            $\ITdelta_{\RM{t,Sc}}$             & [$\si{\micro\meter}$] & 1   & {ReBCO} thickness    \\ 
            $\ITdelta_{\RM{t,Ag}}$             & [$\si{\micro\meter}$] & 2   & Silver thickness    \\
            $\ITdelta_{\RM{t,Cu}}$             & [$\si{\micro\meter}$] & 20  & Copper thickness    \\
            $\ITdelta_{\RM{t,Hs}}$             & [$\si{\micro\meter}$] & 50  & Hastelloy thickness \\
            ${n}$-value                        & [-]                   & 28  & Power-law index     \\
            $\RM{I_{c,min}}(\SI{77}{\kelvin})$  & [$\si{\ampere}$]& 304 & Minimum critical current    \\
            $\RM{I_{c,avg}}(\SI{77}{\kelvin})$  & [$\si{\ampere}$]& 320 & Average critical current\\
            $\ITsigma_{\RM{I_{c}}}$  & [-] & 0.042 & Standard deviation    \\
            \bottomrule
    \end{tabular}           
}

\newcommand{\TABFieldQualityTestCampaign}[0]{
        \begin{tabular}{lccccccl}       
            \toprule
            {No.} & $\SCAl{T}{\RM{op}}$ & $\SCAl{B}{\RM{op}}$    & $\SCAl{r}{\RM{0}}$    & Iron & $\RMDelta\RM{{y}_{l}}$ & $\RMDelta\RM{{y}_{r}}$ & Label \\
            {}    & [$\si{\kelvin}$]    & [$\si{\milli\tesla}$]  & [$\si{\milli\meter}$] &  bars & [$\si{\milli\meter}$]  & [$\si{\milli\meter}$]  & {} \\            
            \midrule 
            \multicolumn{6}{c}{$1^\RM{st}$ HTS holder prototype}\\ 
            \midrule 
            1.  & 300 & 100 & 15 & No & \multicolumn{2}{c}{n.a.} & \multirow{2}{*}{n.a.} \\
            2.  & 77  & 100 & 15 & No & \multicolumn{2}{c}{n.a.}  \\
            3.  & 300 & 100 & 15 & Yes & +5   &  -20 & \multirow{2}{*}{medium} \\
            4.  & 77  & 100 & 15 & Yes & +5   &  -20  \\
            \midrule 
            \multicolumn{6}{c}{$2^\RM{nd}$ HTS holder prototype}\\ 
            \midrule 
            5.  & 300 & 100 & 15 & No & \multicolumn{2}{c}{n.a.} & \multirow{2}{*}{n.a.} \\
            6.  & 77  & 100 & 15 & No & \multicolumn{2}{c}{n.a.}  \\
            7.  & 300 & 100 & 15 & Yes & +5   &  +0  & \multirow{2}{*}{low} \\
            8.  & 77  & 100 & 15 & Yes & +5   &  +0   \\    
            9.  & 300 & 100 & 15 & Yes & +5   &  -20 & \multirow{2}{*}{medium} \\
            10. & 77  & 100 & 15 & Yes & +5   &  -20  \\
            11. & 300 & 100 & 15 & Yes & -20  &  -10 & \multirow{2}{*}{high} \\
            12. & 77  & 100 & 15 & Yes & -20  &  -10  \\
            13. & 300 & 100 & 15 & Yes & -20  &  +5  & \multirow{2}{*}{check} \\
            14. & 77  & 100 & 15 & Yes & -20  &  +5   \\            
            \bottomrule
        \end{tabular}   
        \renewcommand{\arraystretch}{0.8}
}

\newcommand{\TabGeometricalErrorParameters}[0]{
        \begin{tabular}{lllrr}
            \toprule
            {Screen} & Error & Unit & $\RM{1^{st}}$ holder & $\RM{2^{nd}}$ holder \\
            \midrule
            \multirow{3}{*}{Left}  & $\RMepsilon_{\RM{up}}$   & $\si{\micro\meter}$ & $\SI{1200}{}$   & $\SI{-1}{}$    \\
                                   & $\RMepsilon_{\RM{dn}}$  &  $\si{\micro\meter}$ & $\SI{-1900}{}$  & $\SI{170}{}$  \\
                                   & $\RMepsilon_{\RMtheta}$ & $\si{\milli\radian}$ & $\SI{47}{}$     & $\SI{-5}{}$    \\
            \multirow{3}{*}{Right} & $\RMepsilon_{\RM{up}}$  & $\si{\micro\meter}$  & $\SI{-1000}{}$  & $\SI{-75}{}$    \\
                                   & $\RMepsilon_{\RM{dn}}$  & $\si{\micro\meter}$  & $\SI{2100}{}$   & $\SI{5}{}$  \\
                                   & $\RMepsilon_{\RMtheta}$ & $\si{\milli\radian}$ & $\SI{-42}{}$    & $\SI{1}{}$    \\
            \bottomrule            
        \end{tabular}    
}

\newcommand{\TabHALOPerformancResults}[0]{
        \begin{tabular}{lllcccc}
            \toprule
            {No.} & Holder & Field error & $\SCAl{\ITeta}{\RM{g}}$ & $\SCAl{\ITeta}{\RM{m}}$ & $\RM{Q_{g}}$ & $\RM{Q_{m}}$ \\                
            \midrule
            1 & $1^\RM{st}$ & medium  & $0.40$ & $0.55$ & $1.7$ & $2.2$ \\
            2 & $2^\RM{nd}$ & low     & $0.66$ & $0.72$ & $2.9$ & $3.5$ \\          
            3 & $2^\RM{nd}$ & medium  & $0.86$ & $0.68$ & $7.2$ & $3.1$ \\            
            4 & $2^\RM{nd}$ & high    & $0.92$ & $0.74$ & $13.1$ & $3.8$ \\            
            5 & $2^\RM{nd}$ & check   & $0.86$ & $0.66$ & $6.9$ & $2.9$ \\
            \bottomrule
        \end{tabular}    
}

    \usepackage{xr}

    \usepackage{url}
\input{Util/UTIL_MATH_OperatorsV0.1.tex}
\input{Util/UTIL_TEXT_GreekAlphabetV0.1.tex}

\newcommand{\EQdefJpowerLaw}[0]
    { 
    \SCA{\ITrho}(|\VEC{J}|,\VEC{B},\SCA{T}) =
    \frac{\SCAl{E}{\RM{c}}}{\SCAl{J}{\RM{c}}(\VEC{B},\SCA{T})}
    \left(\frac{|\VEC{J}|}{\SCAl{J}{\RM{c}}(\VEC{B},\SCA{T})}\right)^{{n}-1}
    }
    
\newcommand{\EQstrongAampereMaxwell}[0]
    { 
    \CURL\ITmu^{-1}\CURL\VECu{A}{\star} 
    + \SCAu{\ITsigma}{}\dxP{t}\VECu{A}{\star} 
    -\VECl{\ITchi}{\RM{i}}\SCAl{i}{\RM{m}} 
    & = 0
    }
   
\newcommand{\EQstrongHfaraday}[0]
    { 
    \CURL\SCA{\ITrho}\CURL\VEC{H} 
    + \dxP{t}\SCA{\ITmu}\VEC{H} 
    - \CURL\VEClu{\ITchi}{\RM{u}}{k}\SCAlu{u}{\RM{s}}{k} 
    & = 0
    }
   
\newcommand{\EQconstraintIsource}[0]
    { 
    \int\limits_{\SCAlu{\RMOmega}{\RM{H}}{k}}\!\!
    \VEClu{\ITchi}{\RM{u}}{k}\cdot(\CURL\VEC{H})
    \D{\SCA{\RMOmega}} 
    & = \SCAlu{i}{\RM{s}}{k}  
    }

\newcommand{\EQdefMultipoleExpansion}[0]
    { 
    \VEC{B}
    = \frac{\SCAl{B}{1}}{\SI{1e4}{}}
    \sum_{k=1}^{\infty}(\RM{b}_{k}+i\RM{a}_{k})
    \left(\frac{\RM{x}+i\RM{y}}{\SCAl{r}{\RM{0}}}\right)^{k-1}
    }

\newcommand{\EQdefTHDfactor}[0]
    { 
        \SCAl{F}{\RM{d}}
        (\VEC{B}) 
        =
        \sqrt{
        \sum_{k=2}^{\infty}(\RM{a}_{k}^2+\RM{b}_{k}^2)
        }
    }

\newcommand{\EQdefPenaltyFunctionPSO}[0]
    { 
    \begin{aligned}
    \min_{\VEC{\ITvarepsilon}} \quad &
        \sum_{k=2}^{6}
        \left(
        |\SCAl{a}{k,\RM{m}}-\SCAl{a}{k,\RM{s}}(\VEC{\ITvarepsilon})| + 
        |\SCAl{b}{k,\RM{m}}-\SCAl{b}{k,\RM{s}}(\VEC{\ITvarepsilon})|
        \right) \\    
    \textrm{s.t.} \quad & |\SCAl{\ITvarepsilon}{\RM{up}}|,|\SCAl{\ITvarepsilon}{\RM{dn}}|-\SCAl{x}{\RM{c}}\leq  0\\
                        & |\SCAl{\ITvarepsilon}{\RMtheta}|-\RMtheta_{\RM{c}}\leq  0\\ 
    \end{aligned}
    }

\begin{document}


    \title{
    {\Huge
    Proof of Concept of High-Temperature Superconducting Screens\\ for Magnetic Field-Error Cancellation in Accelerator Magnets}}

	\author{
	\normalsize
			L.~Bortot$^{1,2}$,
			M.~Mentink$^{1}$,
			C.~Petrone$^{1}$,
			J.~Van~Nugteren$^{1}$,
			G.~Deferne$^{1}$,
			T.~Koettig$^{1}$,
			G.~Kirby$^{1}$,
			M.~Pentella$^{1,3}$,\\
			J.C.~Perez$^{1}$,
			F.O.~Pincot$^{1}$,
			G.~De~Rijk$^{1}$,
			S.~Russenschuck$^{1}$,
			A.P.~Verweij$^{1}$,
			and~S.~Sch{\"o}ps$^{2}$
		    \vspace{0.25cm}
		}
    \setlength{\affilsep}{0.1cm}
    \affil{\small{$^{1}$ CERN, Espl. des Particules 1, 1211 Geneva, CH}}
    \affil{\small{$^{2}$ Technische Universit{\"a}t Darmstadt, Karolinenplatz 5, 64289 Darmstadt, DE}}
    \affil{\small{$^{3}$ Department of Applied Science and Technology, Polytechnic of Turin, Turin, IT}}  

    \affil{E-mail: \texttt{{lorenzo.bortot@cern.ch}}}

    \date{}    
    \maketitle


%
\begin{abstract}
Accelerators magnets must have minimal magnetic field imperfections for reducing particle-beam instabilities. In the case of coils made of high-temperature superconducting (HTS) tapes, the field imperfections from persistent currents need to be carefully evaluated. In this paper we study the use of superconducting screens based on HTS tapes for reducing the magnetic field imperfections in accelerator magnets. The screens exploit the magnetization by persistent currents to cancel out the magnetic field error. The screens are aligned with the main field components, such that only the undesired field components are compensated. The screens are passive, self-regulating, and do not require any external source of energy. Measurements in liquid nitrogen at $\SI{77}{\kelvin}$ show for dipole-field configurations a significant reduction of the magnetic-field error up to a factor of four. The residual error is explained via numerical simulations, accounting for the geometrical imperfections in the HTS screens, thus achieving satisfactory agreement with experimental results. Simulations show that if screens are increased in width and thickness, and operated at $\SI{4.5}{\kelvin}$, field errors may be eliminated almost entirely for the typical excitation cycles of accelerator magnets.
\end{abstract}


\section{Introduction}    
    \label{SEC_Introduction}
Future circular accelerators for high-energy particle physics are expected to rely on increasingly higher magnetic fields for steering and focusing the particle beams~\cite{fcc2019fcc}. High-temperature superconducting (HTS) tapes based on rare-earth cuprate compounds (ReBCO) have an estimated upper critical field of $\SI{140}{\tesla}$~\cite{golovashkin1991low}, and a critical temperature of $\SI{93}{\kelvin}$. As a consequence, HTS magnets are expected to be operated at fields around 20 T~\cite{van2018toward} and with enthalpy margins one order of magnitude above traditional low temperature superconducting (LTS) materials, such as $\RM{Nb{\HYPHEN}Ti}$ or $\RM{Nb_{3}Sn}$~\cite{wilson1983superconducting}. Therefore, HTS magnets based on ReBCO tapes are a promising technology for high-field magnets in particle accelerators~\cite{van2018toward}.

Accelerator magnets must generate high-quality magnetic fields in their magnet aperture (see e.g.~\cite{bruning2004lhc}), independent of the adopted technology, because field imperfections can lead to particle-beam instabilities~\cite{shi2000collective}. The field quality is determined by magnet design features such as coil geometry and mechanical tolerances, and influenced by material properties such as saturation and hysteresis of the iron yoke. Moreover, time-transient effects such as mechanical deformation due to Lorentz forces, and magnetization due to eddy currents and screening currents in normal conducting and superconducting materials are expected to have detrimental effects. 

Screening currents are particularly relevant in ReBCO tapes, since they behave in the same way as wide, anisotropic mono-filaments, resulting in persistent screening currents. The related magnetization adds an undesired contribution that detrimentally affects the magnetic field quality~\cite{fazilleau2018screening,noguchi2019simple} and decays with a time constant longer than the duty-cycle of the magnet. The degradation of the magnetic field is particularly pronounced at low currents, because the persistent magnetization current is only limited by the superconducting current density, which is the highest at lowest field. 

Previous work for mitigating magnetic field imperfections led to magnetic cloaks for sensors~\cite{gomory2012experimental,tomkow2015combined}, shim coils for magnetic resonance imaging~\cite{frollo2018magnetic} and nuclear magnetic resonance~\cite{wang20163} applications, selective shields for field homogenization in solenoids~\cite{tomkow2019improvement}. 
Recently, persistent-current shims coils were introduced as a conceptual solution for improving the field quality in accelerator magnets~\cite{vannugteren2016persistentshim,van2016high}. The coils are designed as magnet inserts, implementing a canted-cosine theta layout. Moving from this concept, we propose in this paper HTS screens based on superconducting tapes for the passive field-error cancellation in accelerator magnets.

We present a proof-of-concept using HTS screens in a dipole-field configuration. The screens exploit the magnetization produced by persistent currents to shape the magnetic field in the magnet aperture. The screens, made by stacks of tapes arranged in layers, are aligned with the main field component such that only the imperfections in the magnetic field are suppressed. The alignment is made possible thanks to the high aspect ratio between the width and thickness of the tapes. The screens are self-regulating, do not require active control, and store a negligible fraction of the total magnetic energy of the magnet as they do not form any closed loop. The proposed design is called HALO (Harmonics-Absorbing-Layered-Object) which is fully scalable and expandable. 

The prototype for the proof-of-concept consists in two HTS screens aligned in a parallel configuration. The screens are inserted in the aperture of a dipole magnet which magnetic field is perturbed by means of iron bars. Four iron configurations are investigated, differing in the field-error magnitude. The HTS screens are activated by cooling the prototype to $\SI{77}{\kelvin}$ with liquid nitrogen, reducing the magnetic field error by a factor of three to four.
\begin{figure}[tb]
  \centering
	\includegraphics[width=8.5cm]{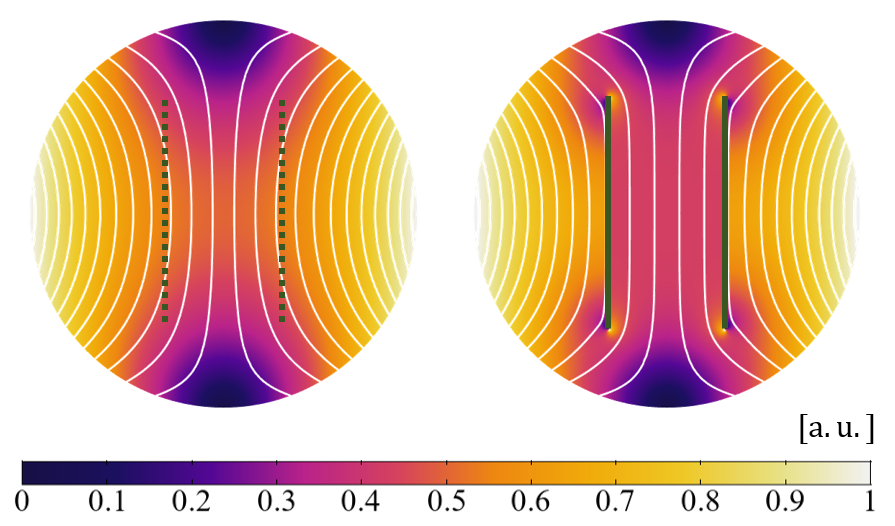}
	\caption{Non-ideal dipole magnetic field, shown before (left) and after (right) the introduction of the HTS screens. The expected and actual position of the screens is marked by dashed and solid lines.}
	\label{FIG_HALO_PRINCIPLE}
\end{figure}

A dedicated 2D numerical model is developed using the finite element method, implementing a coupled $\VEC{A}$-$\VEC{H}$ field formulation~\cite{biro1999edge,brambilla2018finite,dular2019finite} for HTS magnets~\cite{bortot2020coupled}. Simulations are used for evaluating the performance limits of the HTS screens, and tracing the residual error measured in the magnetic field due to geometrical imperfections. Simulations show that by increasing the number of layers and lowering the operational temperature, the HTS screens can be operated in the typical magnetic fields of accelerator magnets. The field-error cancellation can be improved by increasing the width and the number of layers of the HTS screens. The screening current decay is negligible for the operating time of accelerator circuits~\cite{bruning2004lhc}.

The working principle of the HTS screen is discussed in Section~\ref{SEC_WorkingPrinciple}. The experimental setup is shown in Section~\ref{SEC_ExperimentalSetup}, followed by description of the mathematical model in Section~\ref{SEC_MathematicalModel}. Numerical and experimental results are given in Section~\ref{SEC_ExperimentalNumericalResults}, extrapolated in Section~\ref{SEC_Simulations} and discussed in Section~\ref{SEC_Discussion}. Conclusions are given in Section~\ref{SEC_Conclusions}. 
\section{Working Principle} 
    \label{SEC_WorkingPrinciple}
%
\begin{figure}[tb]
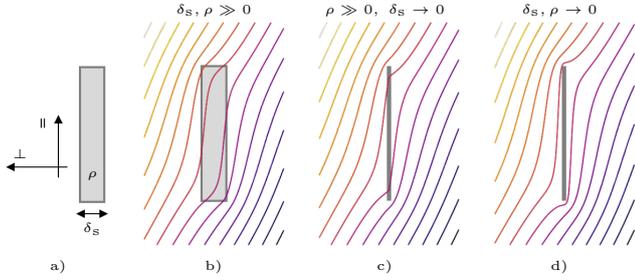

  \centering
    \FigWorkingPrinciple
	\caption{
	a) Cross-section of a non-magnetized, conducting shell of thickness $\SCAl{\ITdelta}{\RM{s}}$ and resistivity $\ITrho$. The magnetic flux lines distribution is shown for the cases of b) finite $\SCAl{\ITdelta}{\RM{s}}$ and $\ITrho$, c) negligible $\SCAl{\ITdelta}{\RM{s}}$ but finite $\ITrho$, and d) negligible $\SCAl{\ITdelta}{\RM{s}}$ and $\ITrho$.}
	\label{FIG_WorkingPrinciple}
\end{figure}
The working principle is discussed with regards to a non-magnetized shell with finite thickness $\SCAl{\ITdelta}{\RM{s}}$ and constant resistivity $\ITrho$. The cross section of the shell is shown in in Fig.~\ref{FIG_WorkingPrinciple}a, together with a local coordinate system $(\PARA,\PERP)$ oriented accordingly to the shell wide surface. The shell is exposed to an externally applied magnetic flux density $\VECl{B}{\RM{s}}(\bm{r},t)$ depending on space $\bm{r}\in\mathbb{R}^3$ and time  $\SCA{t}\in\mathbb{R}^3$, with initial condition $\VECl{B}{\RM{s}}(\bm{r},0)=0$.
All the magnetic properties are considered as constant. Starting from Faraday's law, the screening current density $\VECl{J}{\RM{i}}$ induced in the shell is obtained as 
    \begin{align}
        \label{EQ_faraday}
            \CURL\ITrho\VECl{J}{\RM{i}}
            +\dxP{t}(\VECl{B}{\RM{s}}+\VECl{B}{\RM{i}})
            =0,      
    \end{align}
where $\VEC{E}=\ITrho\VECl{J}{\RM{i}}$ denotes the electric field driving the screening currents, and $\VECl{B}{\RM{i}}$ is the magnetic contribution from the shell to the magnetic flux density $\VEC{B}=\VECl{B}{\RM{s}}+\VECl{B}{\RM{i}}$. The distribution of the magnetic flux lines is shown in Fig.~\ref{FIG_WorkingPrinciple}b.

\subsection{Ideal Screens} 
    \label{SubSEC_IdealScreens}
By assuming negligible thickness for the shell, i.e. $\SCAl{\ITdelta}{\RM{s}}\to0$ (see Fig.~\ref{FIG_WorkingPrinciple}c), the magnetic coupling occurs only for the normal component of the external field source, leaving the parallel component unaffected. At the same time, the induced current density flows only in the plane of the shell. A local coordinate system is conveniently oriented accordingly to the wide surface of the shell, such that the differential operators and vectors $\VEC{v}$ are decomposed into their tangential (superscript $\parallel$) and normal (superscript $\perp$) components, that is, $\GRAD=\GRAD^{\PARA}+\GRAD^{\PERP}$ and $\VEC{v}=\VECu{v}{\PARA}+\VECu{v}{\PERP}$. With these definitions,~\eqref{EQ_faraday} is reduced to
    \begin{align}
        \label{EQ_faradayShell}
            \GRAD^{\PERP}\ITrho\VEClu{J}{\RM{i}}{\PARA}
            +\dxP{t}(\VEClu{B}{\RM{s}}{\PERP}+\VEClu{B}{\RM{i}}{\PERP})
            =0.     
    \end{align}
By assuming negligible resistivity for the shell, $\ITrho\to0$ (see Fig.~\ref{FIG_WorkingPrinciple}d), the material becomes a perfect electrical conductor (PEC). As a consequence, \eqref{EQ_faradayShell} is reduced to ${\dxP{t}(\VEClu{B}{\RM{s}}{\PERP}+\VEClu{B}{\RM{i}}{\PERP})=0}$ showing that the magnetic field remains fixed within the shell. 
Since the external field was initially zero, $\VEClu{B}{\RM{i}}{\PERP}$ is always equal and opposite to $\VEClu{B}{\RM{s}}{\PERP}$. The induced currents become persistent, exhibiting no decay time, and their magnitude is determined by the magnetization required to completely cancel out the externally applied field. PECs are therefore ideal magnetic screens.

\subsection{Practical Screens} 
    \label{SUBSEC_PracticalScreens}
%
\begin{figure}[tb]
    \centering
    \includegraphics[width=8.5cm]{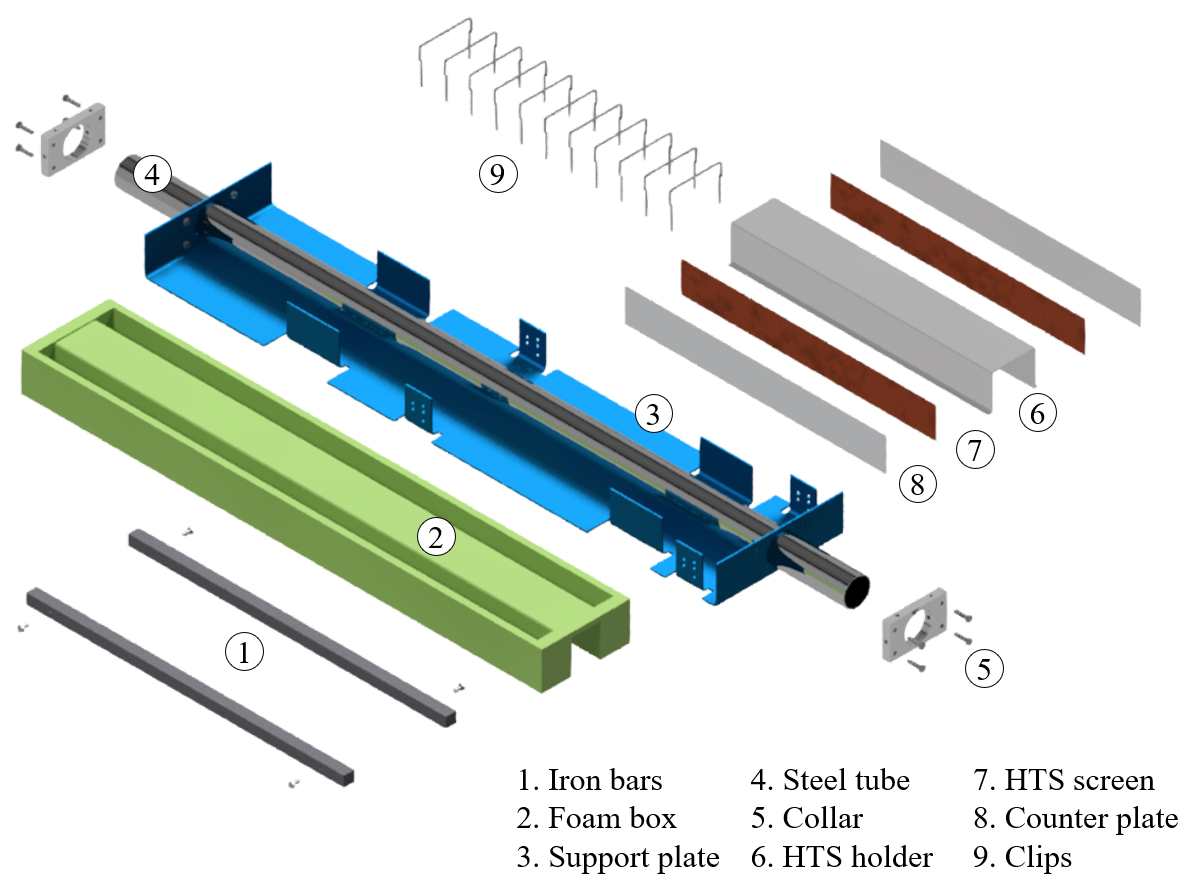}
    \caption{Exploded view of the parts of the experimental assembly. 1. iron bars, 2. polymide foam box, 3. aluminum support plate, 4. stainless steel tube, 5. aluminum collars, 6. aluminum HTS holder, 7. HTS screens, 8. aluminum counter plates, 9. stainless steel clips.}
	\label{FIG_SETUP_MODEL_01}    
\end{figure}
\begin{figure}[tb]
    \centering
    \includegraphics[width=8.5cm]{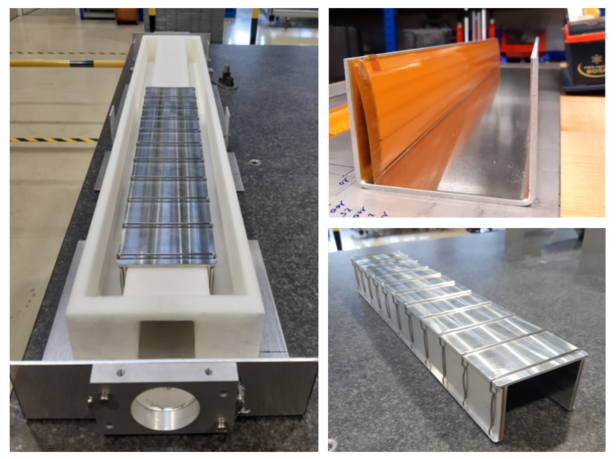}
    \caption{Left: experimental assembly. The support plate at the bottom bottom holds the iron bars, the foam box and the shell which is blocked by means of two supports at the ends. The holder is settled inside the foam box. Right, top: first HTS holder prototype, obtained by bending an aluminum plate and fixing the screen using adhesive tape. Right, bottom: second HTS holder prototype, obtained by machining an aluminum block and fixing the tapes using aluminum counter-plates and clips.}
	\label{FIG_SETUP_MODEL_02}    
\end{figure}

Although perfect electrical conductors are a mathematical abstraction, the superconducting properties of the tapes provide a reasonable approximation of $\ITrho\to0$, allowing for persistent magnetization generated by screening currents. This phenomenon is combined together with the strong geometrical anisotropy of the tapes and their negligible thickness, creating a selective magnetic coupling with respect to the spatial components of the applied field. 

By choosing a suitable orientation, the tapes can be used to "guide" the magnetic field, achieving a field correction only for specific field components. This is obtained by aligning the tapes with the the main field direction, such that the cancellation effect occurs only for the undesired field components. The tapes can be arranged side by side into layers, increasing the equivalent screening surface. Layers can be stacked on top of each other to increase the magnetic screening properties. The working principle is illustrated by the example provided in Fig.~\ref{FIG_HALO_PRINCIPLE}, where a non-ideal dipole magnetic field is shown before (left) and after (right) the introduction of the HTS screens.

The working principle is applicable also to 2D magnetic field configurations with higher number of magnetic poles (e.g. quadrupole fields) as long as the superconducting screens are shaped accordingly to the main field component.

\section{Experimental Setup}
    \label{SEC_ExperimentalSetup}
The proof of concept aims at demonstrating that the magnetic field quality can be improved in a given region of space by means of persistent screening currents. The proof is achieved by using differential measurements, by assessing the magnetic field quality with and without the presence of the HTS screens. 

Four key-elements are included in the experimental setup: 1) a dipole field of known magnetic properties, provided by the reference dipole MCB24 from the Magnetic Measurement laboratory at {CERN}; 2) a source of field perturbation, that is, two iron bars introduced in the magnet aperture; 3) the field-error cancellation source, provided by means of two HTS screens; 4) a magnetic measurement system for characterizing the field quality, composed of a rotating induction coil~\cite{walckiers1992harmonic,jain1998harmonic}, a motor drive, and the DAQ system for processing the probe signal. The experimental setup is detailed in the remainder of the section. 

\subsection{Mechanical Assembly}
    \label{SUBSEC_MechanicalAssembly}
The mechanical assembly is composed of an aluminum base plate, a stainless steel shell-tube, two iron bars made of pure iron and a box made of polymide foam, hosting the HTS holder with the superconducting screens. The assembly is shown in Fig.~\ref{FIG_SETUP_MODEL_01} and Fig.~\ref{FIG_SETUP_MODEL_02} (left). 

The plate provides both the mechanical reference for the alignment in the magnet aperture and the mechanical support for the remaining parts. The shell-tube is bolted by means of collars to the front and back fins of the plate. The iron bars are bolted on perforated side fins, at each side of the shell. A set of holes allows the iron bars to be vertically displaced, in order to investigate the field error cancellation for different field-error scenarios. 

The box is leak-tight and works as a cryostat for the cool down of the HTS screens to $\SI{77}{\kelvin}$ in a bath of liquid nitrogen. The box is machined such that a central groove on the bottom ensures the clearance for the shell containing the rotating coil probe, whereas two lateral grooves on the top allow the HTS-screen holder to slide into the box. The HTS-screen holder provides the mechanical support for keeping the screens in parallel position. For this setup, two HTS-screen holder prototypes were developed and tested. 

\subsection{HTS Screens}
    \label{SUBSEC_HTSScreens}
%
\begin{figure}[tb]
    \centering
    \includegraphics[width=8.5cm]{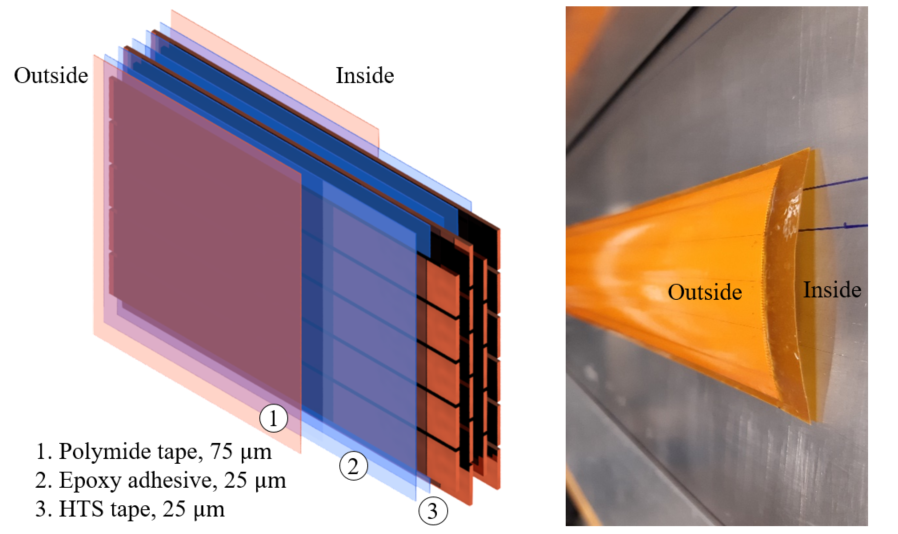};   
    \caption{Left: multi-layered composite structure characterizing the HTS screens. Four layers of HTS tapes are arranged in a stack by means of adhesive tape which provides also electrical insulation. Right: one of the two HTS screens used for the proof of concept.}
	\label{FIG_HALOarchitecture}    
\end{figure}
\begin{table}[tb]
    \centering
    \caption{Tape specifications (self field)}
    \label{TAB_TapeSpecifications}
    \TabTapeSpecifications
\end{table}

The HTS screens are made of commercially available, second-generation HTS tapes; their relevant parameters are given in Table~\ref{TAB_TapeSpecifications}. The HTS spool is cut in $\SI{500}{\milli\meter}$-long tapes which are arranged in a multi-layered composite structure, as shown in Fig.~\ref{FIG_HALOarchitecture}, left. Each layer is obtained by bonding the tapes along their narrow edge over a $\SI{25}{\micro\meter}$-thick layer of epoxy glue, and then the layers are stacked on top of each other. 

An offset equal to half of the tape width is introduced between each layer, leading to a brick wall structure. The offset improves the screening properties of the structure by preventing the magnetic flux lines to penetrate in the gaps between the tapes. For symmetry reasons, the even-order layers are one tape-width less wide than the odd-order layers. The screens are sealed with $\SI{25}{\micro\meter}$ polymide foils, applied on each side. Such structure is mechanically flexible and scalable for both the screen width and thickness, ensuring electrical insulation between each HTS layer and for the overall screen. 

Two four-layer, $\SI{60}{\milli\meter}$-wide screens are manufactured for the proof of concept, with the layers containing 5 and 4 tapes. The maximum width of the screens is limited by the available space within the aperture of the reference magnet. One of the screens is shown in Fig.~\ref{FIG_HALOarchitecture}, right. The screen curvature is a consequence of stacking all the tapes on the same side, causing an amplification of the typical tape convexity. This behavior is not an issue as the screens are flexible enough to be straightened by the HTS holder. Moreover, the curvature may be mitigated by flipping upside-down the tapes in every second layer of each screen.
  
\subsection{HTS Holder - First Prototype}
    \label{SUBSEC_HTSHolderFirstPrototype}
The first prototype aimed for the simplicity of construction (see Fig.~\ref{FIG_SETUP_MODEL_02}, top right). The holder was obtained by bending an aluminum plate, and is used to keep in place the screens by compressing them between the holder and the foam box. It was subsequently found that the mechanical alignment of the first prototype is not sufficiently accurate for the proof of concept. The bending process introduced internal stresses in the material which were released during the cool-down, degrading the mechanical alignment of the screens. 

\subsection{HTS Holder - Second Prototype}
    \label{SUBSEC_HTSHolderSecondPrototype}
The experience gained from the first holder led to the improved design of the second prototype (see Fig.~\ref{FIG_SETUP_MODEL_02}, bottom right), where a compression force keeps the screens straightened by means of two aluminum counter-plates pushed by clips made of non-magnetic stainless steel, positioned at every $\SI{50}{\milli\meter}$ along the holder. With respect to the previous version, the second prototype is characterized by a lower mechanical tolerance and higher stiffness, and was therefore used for the proof of concept.

\subsection{Reference Dipole}
    \label{SUBSEC_ReferenceDipole}
%
\begin{figure}[tb]
    \centering
    \includegraphics[width=8.5cm]{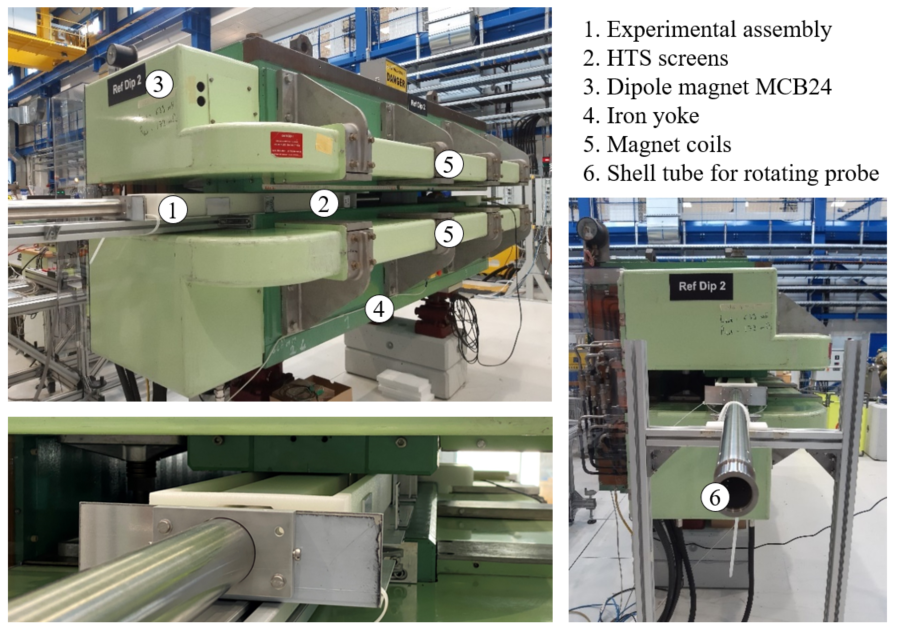}
    \caption{Top Left: lateral view of the reference dipole magnet MCB24, with the setup mounted in the magnet aperture. Bottom left: detail of the magnet aperture. Right: front view of the magnet.}
	\label{FIG_SETUP_MAGNET_2}    
\end{figure}
\begin{figure}[tb]
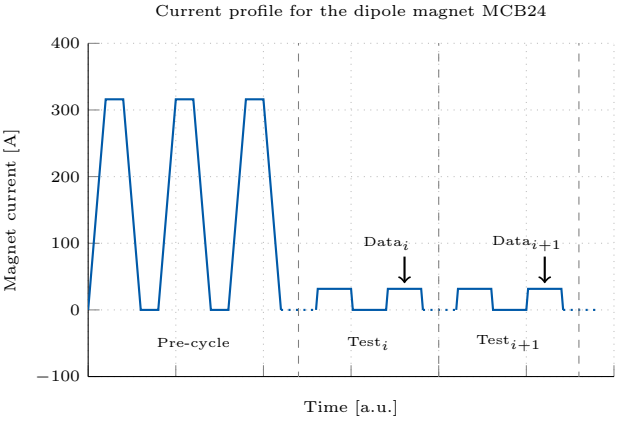

    \centering
    \def\fileName{Content/Data/Data_various.csv}
    \FigLineMagnetCurrentProfile{\fileName}         
    \caption{Current cycle used in the reference dipole magnet MCB24 for the experimental campaign.}
    \label{FIG_MagnetCurrentProfile}
\end{figure}

The setup was fixed into the aperture of the reference dipole magnet MCB24 in the magnetic measurement lab at {CERN} (see Fig.~\ref{FIG_SETUP_MAGNET_2}). As the foam box was designed $\SI{500}{\milli\meter}$ longer than the screen, it stuck out from the magnet aperture (see Fig.~\ref{FIG_SETUP_MAGNET_2}, right), leaving sufficient space for adding the coolant from outside the magnet coils. 

The peak dipole field for the proof-of-concept is $\SI{100}{\milli\tesla}$, to stay below the penetration field of the tapes and to avoid excessive magnetic forces on the iron bars. The magnet is normal-conducting and has a linear transfer function of $\SI{316}{\ampere\per\tesla}$ for the $\SI{0}{}$-$\SI{1}{\tesla}$ field range. The magnet is operated subsequently in a de-gaussing cycle. The profile used for the source current $\SCAl{i}{\RM{m}}(t)$ is shown in Fig.~\ref{FIG_MagnetCurrentProfile}. The current follows first a trapezoidal pre-cycle, going from  from zero up to the nominal current $\SCAl{I}{\RM{n}}=\SI{316}{\ampere}$, and back. Then, each test cycle is composed by two consecutive trapezoidal curves, up to the peak source current $\SCAl{I}{\RM{s}}=\SI{31.6}{\ampere}$, delivering a magnetic field in the magnet aperture of $\SI{100}{\milli\tesla}$. A constant current is retained for about $\SI{120}{\second}$, to settle any dynamic effects. The measurements in each cycle are acquired for both the current plateaus, showing negligible difference. The data presented in this work are always taken from the second curve, where the HTS tapes are already magnetized. 
 
\subsection{Measurement System}
    \label{SUBSEC_MechanicalSupport}
Rotating-coil magnetometers, also known as harmonic coils, are electromagnetic transducers for measuring the $\SCAl{B}{k}$ and $\SCAl{A}{k}$ field multipoles. The coil shaft is positioned parallel to the magnetic axis of the magnet, and rotated in the magnet aperture. The change of flux linkage $\RMPhi$ induces, by integral Faraday\'{}s law $\SCAl{U}{\RM{m}} = -\RM{d}_{t}{\RMPhi}$, a voltage signal $\SCAl{U}{\RM{m}}$ which is measured at the terminals of the coil. By integrating in time the voltage signal, the flux linkage is obtained and given as a function of the series expansion of the radial field~\cite{russenschuck2011field}. 

The rotating coil used for the HALO characterization is composed of a Printed-Circuit Board (PCB), ($\SI{36.5}{\milli\meter}$ in length and~$\SI{47}{\milli\meter}$ in width), aligned with the longitudinal center of the HTS screens. The PCB contains five coils mounted radially, with an  active surface of $\SI{0.03186}{\meter\squared}$. {CERN} proprietary digital cards~\cite{arpaia2012performance} integrate the induced voltages in the coils rotating at a frequency of~$\SI{1}{Hz}$. Each measurement is given by the average of sixty rotations of the coils. The measurement results delivers a typical precision of a magnetic-field harmonic of $\pm0.05$ units. 
\section{Mathematical Model}
    \label{SEC_MathematicalModel}
%
\begin{figure}[tb]
    \centering
    \includegraphics[width=8.5cm]{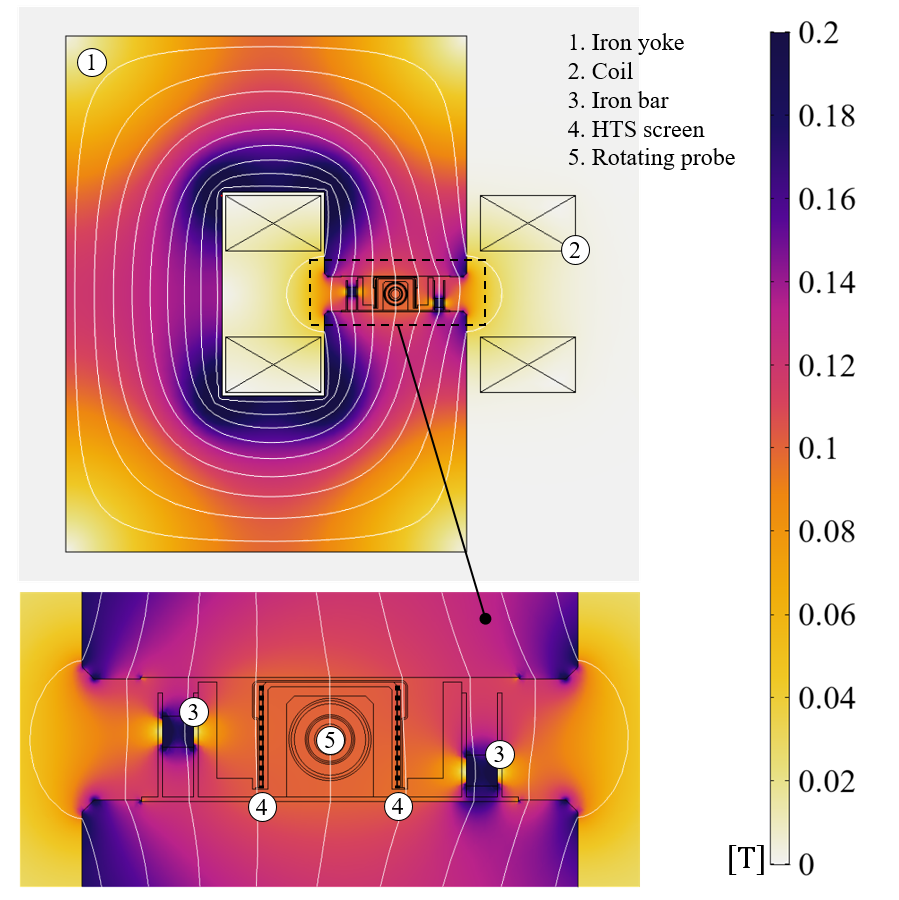}
    \caption{Top: magnetic field, in tesla, in the cross section of the reference dipole magnet. The field source is provided by means of a normal conducting coil, marked with crossed domains. The c-shaped iron yoke guides the field lines in the magnet aperture, where the experimental setup is positioned. Bottom: detailed view of the magnet aperture containing the experimental setup and the iron bars which introduce a field perturbation. The position of the HTS screens is highlighted by dashed lines.}
	\label{FIG_FieldMagnet}    
\end{figure}

This field problem is formulated by combining a domain decomposition strategy with a dedicated coupled field formulation derived from~\cite{bortot2020coupled}. The computational domain $\SCA{\RMOmega}$ containing the reference dipole magnet and the experimental setup is presented in Fig.~\ref{FIG_FieldMagnet}. The magnetic field, given in tesla, in the cross section of the reference dipole magnet is shown on the top. A detailed view of the magnet aperture containing the experimental setup and the iron bars which introduce a field perturbation is given on the bottom. The position of the HTS screens is highlighted by dashed lines.

The domain is decomposed into the regions $\SCAl{\RMOmega}{\RM{H}}$ and $\SCAl{\RMOmega}{\RM{A}}$, oriented with the unit vector $\VEC{n}$, such that $\SCAl{\BAR{\RMOmega}}{\RM{H}}\cup\SCAl{\BAR{\RMOmega}}{\RM{A}}=\SCA{\BAR{\RMOmega}}$. The region $\SCAl{\RMOmega}{\RM{H}}$, corresponding to the HTS screens and containing both superconducting and normal conducting materials, is further subdivided into $\SCAlu{{\RMOmega}}{\RM{H}}{k}$ regions, each representing one tape, such that $\SCAl{\BAR{\RMOmega}}{\RM{H}}=\sum_{k=1}^{\RM{N_t}}\SCAlu{\BAR{\RMOmega}}{\RM{H}}{k}$ where $\RM{N_t}$ is the number of tapes. The region $\SCAl{\RMOmega}{\RM{A}}$ contains both normal-conducting and non-conducting materials and it is given by the remainder of the magnet such as the mechanical structure of the experimental setup, the iron bars, the iron yoke, the normal conducting coils, and the air region.

A constant magnetic permeability $\ITmu$ is assumed in $\SCAl{\RMOmega}{\RM{H}}$, whereas a nonlinear dependency from the magnetic field $\VEC{B}$ is considered for the iron yoke and the iron bars in $\SCAl{\RMOmega}{\RM{A}}$, as $\ITmu(\VEC{B})$. 

The domains $\SCAl{\RMOmega}{\RM{A}}$ and $\SCAl{\RMOmega}{\RM{H}}$ are equipped with the winding  functions~\cite{rodriguez2008voltage,schops2013winding} $\VECl{\ITchi}{\RM{i}}$ and $\VEClu{\ITchi}{\RM{u}}{k}$, with $k=1,..,\RM{N_t}$. The first represents a current distribution function for stranded conductors assigned to the magnet coil, whereas the second contains $\RM{N_t}$ voltage distribution functions, one for each of the $k$-th HTS tapes in the HTS screens.

The field problem is solved under magnetoquasistatic assumptions for the reduced magnetic vector potential $\VECu{A}{\star}$~\cite{emson1983optimal} in $\SCAl{\RMOmega}{\RM{A}}$, and for the magnetic field strength $\VEC{H}$~\cite{bossavit1988rationale,brambilla2006development} in $\SCAl{\RMOmega}{\RM{H}}$, with $\VECu{A}{\star}\times\VEC{n}=0$ as Neumann boundary condition on the exterior boundary. The formulation reads for $k=1,..,\RM{N_t}$
    \begin{align}
        \label{EQstrongAampereMaxwell}
            \EQstrongAampereMaxwell{}\ \text{in}\ \SCAl{\RMOmega}{\RM{A}}, \\
        \label{EQstrongHfaraday}
            \EQstrongHfaraday{}\ \text{in}\ \SCAlu{\RMOmega}{\RM{H}}{k}, \\ 
        \label{EQconstraintIsource}
            \EQconstraintIsource{},
    \end{align}
where $\SCA{\ITsigma}$ and $\SCA{\ITrho}$ are the electrical conductivity and resistivity, $\SCAl{i}{\RM{m}}$ is the source current in the dipole magnet, $\SCAlu{u}{\RM{s}}{k}$ is source voltage for the $k$-th tape treated as an algebraic unknown, and $\VEClu{i}{\RM{s}}{k}$ is the source current for $k$-th tape imposed via a constraint equation (i.e., a Lagrange multiplier). For the purpose of this analysis, $\SCAlu{i}{\RM{s}}{k}=0\ {\forall}{k} \in [1,\RM{N_t}]$, as the tapes are passive and do not form any closed loops.

The fields $\VECu{A}{\star}$ and $\VEC{H}$ are linked via continuity conditions at the interface of the domains $\SCAl{\RMGamma}{\RM{HA}}$, given by composition of the boundaries of all the HTS tapes. In particular, the continuity of the normal component of the magnetic flux density $\VECl{B}{\RM{n}}$ and the current density $\VECl{J}{\RM{n}}$, and the tangential component of the magnetic field $\VECl{H}{\RM{t}}$ and electric field $\VECl{E}{\RM{t}}$ are imposed, ensuring the consistency of the overall field solution.

The field formulation proposed in~(\ref{EQstrongAampereMaxwell}-\ref{EQconstraintIsource}) avoids the use of electrical conductivity for the superconducting domains ($\ITsigma\to\infty$) and the electrical resistivity for the non-conducting domains ($\ITrho\to\infty$), such that the material properties remain finite~\cite{ruiz2004numerical,dular2019finite}. 

\subsection{Constitutive relations}
    \label{SUBSEC_ConstitutiveEquation}
\begin{figure}[tb]
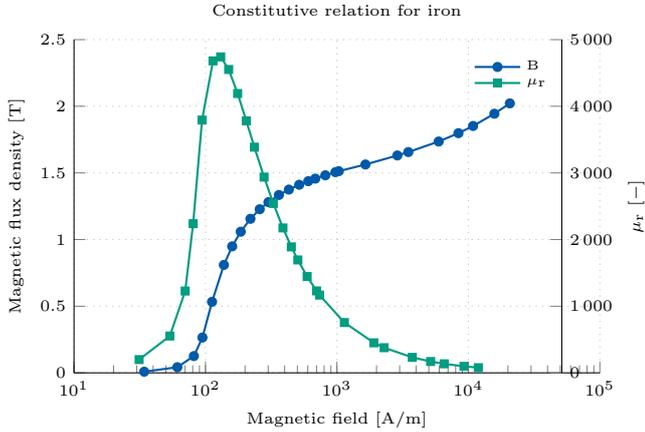

    \centering
    \def\fileName{Content/Data/Data_various.csv}
    \FigBHcurvePureIron{\fileName} 
    \vskip -0.35cm
    \caption{Measured B-H curve for the nonlinear $\ITmu(\VEC{B})$ relation.}
    \label{FIG_BHcurvePureIron}
\end{figure}

The ferromagnetic materials included in the model are made of pure iron ($\RM{Fe}>99.8\%$). The B-H relation was measured for a sample~\cite{pentella2021pureiron} and the resulting curve, displayed in Fig.~\ref{FIG_BHcurvePureIron}, is used in the model for the $\ITmu(\VEC{B})$ relation.

The highly nonlinear electric field / current density relation~(e.g.~\cite{wu1987superconductivity}) characterizing HTS materials is modeled by means of a power law~\cite{rhyner1993magnetic}. Such simplified relation for the resistivity neglects frozen-field phenomena, occurring only in the low current density regime. The power law provides faster field relaxation and decay rates for screening-current phenomena with respect to the more complex percolation-depinning law~\cite{yamafuji1997current,sirois2018comparison}, therefore it is retained as conservative assumption. The resistivity is given by
    \begin{align}
        \label{EQ_powerLaw}
            \EQdefJpowerLaw{},      
    \end{align}
where $\VEC{J}$ is the current density, $\SCAl{E}{\RM{c}}$ is the critical electric field, set to $\SI{1e-4}{\volt\per\meter}$~\cite{dew1988model}, and the material- and field-dependent parameters $\SCAl{J}{\RM{c}}$ and ${n}$ are the anisotropic critical current density and the power-law index, respectively. 

\begin{figure}[tb]
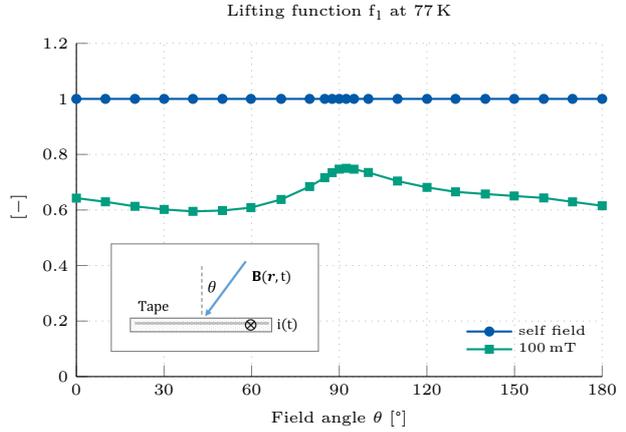

    \centering
    \def\fileName{Content/Data/Data_various.csv}
    \FigLiftingFunction{\fileName}         
    \caption{Lifting function as a function of the field angle, for tapes at $\SI{77}{\kelvin}$, parametrized with the magnetic field magnitude.}
    \label{FIG_LiftingFunction}
\end{figure}

The lifting function $\SCAl{f}{\RM{l}}(\VEC{B},\SCA{T})$ implemented for $\SCAl{J}{\RM{c}}$ is shown in Fig.~\ref{FIG_LiftingFunction}, for a background field of $\SI{100}{\milli\tesla}$. Data are taken from~\cite{hu2016modeling}, where tapes from the same producer and technology were characterized. Then, $\SCAl{J}{\RM{c}}$ is obtained from $\SCAl{J}{\RM{c}}=\SCAl{f}{\RM{l}}\RM{I_{c,min}}/\SCAl{S}{\RM{Sc}}$ where the minimum critical current and the superconductor cross section $\SCAl{S}{\RM{Sc}}$ are taken from Table~\ref{TAB_TapeSpecifications}.

\subsection{Magnetic Field Quality}
    \label{SUBSEC_MagneticFieldQuality}
In accelerator magnets, the magnetic field quality is traditionally defined as a set of Fourier coefficients known as multipoles. These field coefficients are given by the solution of the Laplace equation $\LAPL\VEC{A}=0$ in the magnet aperture, for the cross sectional plane of the magnet aperture. 

The multipole expansion series ~\cite{russenschuck2011field} is calculated at the reference radius $\SCAl{r}{\RM{0}}$, usually chosen as 2/3 of the magnet aperture. By decomposing the magnetic field $\VEC{B}$ by means of complex notation as $\VEC{B}=\SCAl{B}{\RM{y}}+i\SCAl{B}{\RM{x}}$, the series reads
    \begin{align}
        \label{EQ_multipoleExpansion}
            \EQdefMultipoleExpansion{},      
    \end{align}
where $\SCAl{B}{1}$ is the dipole field component, $k$ is the order of the eigensolution generated by ideal magnet geometries, and $\RM{b}_{k}$ and $\RM{a}_{k}$ are the 2$k$-pole normal and skew coefficients, given in units ($\SI{1e-4}{}$ of the main field). Therefore, in dipole fields the magnetic field error is quantified by the magnitude of the $k\geq2$ multipole coefficients.

The total harmonic distortion (THD) factor $\SCAl{F}{\RM{d}}$ is a scalar quantity defined for $\SCA{r}=\SCAl{r}{\RM{0}}$ as 
    \begin{align}
        \label{EQ_THDfactor}
            \EQdefTHDfactor{}.     
    \end{align}
In this paper, the calculation of $\SCAl{F}{\RM{d}}$ is done up to the dodecapole components ($k=6$), considering that higher-order field distortions are found to be generally of much lower magnitude. The numerical evaluation of the field quality is obtained by sampling the magnetic field solution along the reference circumference over 4096 points homogeneously distributed. Subsequently, the multipole coefficients are calculated by means of a Fast Fourier Transform algorithm applied to the radial field component~\cite{russenschuck2011field}. 

\subsection{Geometrical Quality}
    \label{SUBSEC_GeometricalQuality}
%
\begin{figure}[tb]
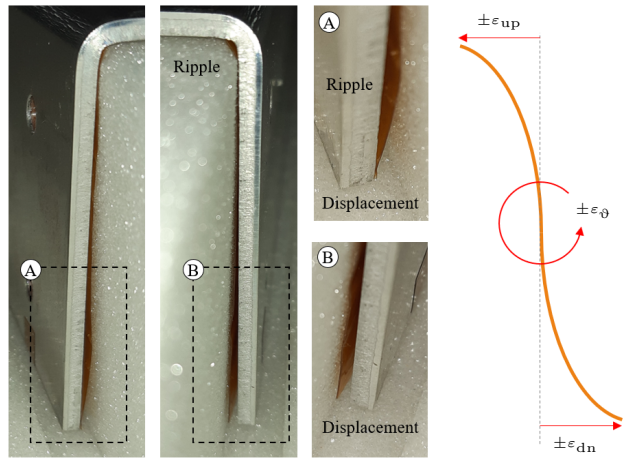

  \centering
    \figHALOVisualInspection
	\caption{Left: visual inspection of the first HTS holder prototype, highlighting gaps between the holder and the foam box, as well as displacements and corrugations for both the left and right HTS screens. Right: geometrical error parameters, used to implement a realistic geometry for the HTS screen.}
	\label{FIG_HALOVisualInspection}
\end{figure}
A visual inspection of the experimental assembly highlighted gaps between the first prototype of the HTS holder and the foam box, as well as displacements and corrugations for both the left and right screens; see Fig.~\ref{FIG_HALOVisualInspection}, left. Such undesired geometrical errors $\VEC{\ITvarepsilon}$ arise from intrinsic stresses in the HTS tapes and were found to detrimentally affect the field quality.

The geometrical errors introduced in the model are shown for one screen in Fig.~\ref{FIG_HALOVisualInspection}, right. 
The HTS screens are implemented as two joint arcs of a parabola, whose shape is determined by the parameters $\ITvarepsilon_{\RM{up}}$ and $\ITvarepsilon_{\RM{dn}}$. Screens are also allowed to rotate around their central point, accordingly to the angle $\ITvarepsilon_{\RMtheta}$. 
Three degrees of freedom are introduced for each screen, therefore the geometrical imperfection is defined for this model as
$\VEC{\ITvarepsilon}=\left[\SCAlu{\ITvarepsilon}{\RM{up}}{\RM{l}},\SCAlu{\ITvarepsilon}{\RM{up}}{\RM{r}},\SCAlu{\ITvarepsilon}{\RM{dn}}{\RM{l}},\SCAlu{\ITvarepsilon}{\RM{dn}}{\RM{r}},\SCAlu{\ITvarepsilon}{\RMtheta}{\RM{l}},\SCAlu{\ITvarepsilon}{\RMtheta}{\RM{r}}\right]$, with the superscripts $l$ and $r$ referring to the left and right screens. It is worth noting that flat screens are still possible with $\ITvarepsilon_{\RM{up}},\ITvarepsilon_{\RM{dn}}\to0$, whereas the parallel condition requires also $\ITvarepsilon_{\RMtheta}\to0$.

The \textsc{Matlab}$^{\circledR}$ ~\cite{matlab2018b} implementation of the particle swarm optimization (PSO)~\cite{kennedy1995particle} is used to determine the parameters. The penalty function adopted for the PSO minimizes the difference between field quality measurements and simulations. The optimization problem reads
\begin{equation}
        \label{EQ_defPenaltyFunctionPSO} 
        \EQdefPenaltyFunctionPSO 
\end{equation}
with $\SCAl{a}{k,\RM{m}}$ and $\SCAl{b}{k,\RM{m}}$ derived from measurements, whereas $\SCAl{a}{k,\RM{s}}(\VEC{\ITvarepsilon})$ and $\SCAl{b}{k,\RM{s}}(\VEC{\ITvarepsilon})$ are calculated numerically. The boundary constraints are set to $\SCAl{x}{\RM{c}}=\SI{2.5}{mm}$ and $\RMtheta_{\RM{c}}=\SI{50}{\milli\radian}$. The index $k$ is limited to the dodecapole component, consistent with the definition of the THD factor. As discussed in section~\ref{SEC_ExperimentalNumericalResults}, the simulated field error arising from geometrical imperfection is matched with the experimental observations.
\section{Experimental and Numerical Results} 
    \label{SEC_ExperimentalNumericalResults}
%
\begin{table}[tb]
    \caption{Test campaign}
    \label{TAB_FieldQualityTestCampaign}
    \centering
    \TABFieldQualityTestCampaign
\end{table}
The test campaign for the HALO experiments is organized in two parts, assessing the behavior of the HALO device without and with the iron bars.

In the first part, the HTS screens are characterized both at room temperature ($\SI{300}{\kelvin}$) and in liquid nitrogen ($\SI{77}{\kelvin}$), in a dipole background field. The first measurement determines the magnetic contribution from the normal conducting materials in the assembly. The second measurement quantifies the magnetic coupling of the HTS screens with the background field, which is strongly influenced by the precise alignment of the tapes with respect to the magnetic field lines. The two tests are carried out for both the prototypes of the HTS holder. Subsequently, these results are used for fitting the geometrical error parameters and, therefore, calibrating the numerical model.

In the second part, iron bars are introduced in the magnet aperture, adding a field error to the dipole field of the reference magnet. The field error is modulated by applying to the left and right iron bars the vertical offsets $\RMDelta\RM{{y}_{l}}$ and $\RMDelta\RM{{y}_{r}}$, in a range between $+5$ and $-20$ $\si{\milli\meter}$ with respect to the horizontal mid-plane of the rotating coil probe. Four different scenarios are proposed. The first three scenarios feature an increasing magnitude for the field error, therefore they are labeled as $low$, $medium$ and $high$. The fourth scenario provided a sanity check as its configuration is anti-symmetric with respect to the second scenario, therefore it is labeled as $check$. For each scenario, the field quality is measured first at $\SI{300}{\kelvin}$, being affected only by the iron, and then at $\SI{77}{\kelvin}$, where also the HTS screens are active. In this way, a comparison of the two measurements gives the net contribution of the HTS screens to the field-error cancellation.

The most relevant features for each of the tests are summarized in Table~\ref{TAB_FieldQualityTestCampaign}. The first HTS holder prototype is characterized only for the $medium$~error scenario, whereas the second one is characterized for all scenarios. All the test are performed in a $\SI{100}{\milli\tesla}$ background dipole field, determined by evaluating the Lorentz forces acting on the iron bars. All the multipoles are evaluated at a reference radius of $\SCAl{r}{\RM{0}}=\SI{15}{\milli\meter}$.

Measurements are compared with simulations. With respect to the tests at $\SI{77}{\kelvin}$, two geometrical models for each HTS screen prototype are considered. The first assumes perfectly parallel and flat HTS tapes, whereas the second includes the geometrical errors introduced in section~\ref{SUBSEC_GeometricalQuality}. In the following, simulation results are labeled as flat and curved, referring to the geometry adopted for the screens. The flat model allows calculating the upper limit for the screen performance, whereas the curved model reproduces the behavior of the experimental setup. The calibration of the second model is discussed in the next section. 

All the simulations are carried out on a standard workstation (Intel$^{\circledR}$ Core i7-3770 {CPU} $@$ $\SI{3.40}{\giga\hertz}$, $\SI{32}{\giga\byte}$ of {RAM}, Windows-$10^{\circledR}$ Enterprise 64-bit operating system), using the proprietary {FEM} solver \textsc{{COMSOL} {M}ultiphysics}$^{\circledR}$~\cite{comsol2005comsol}.

\subsection{Geometrical Quality Analysis}
    \label{SUBSEC_GeometricalQualityAnalysis} 
%
\begin{table}[tb]
    \caption{Geometrical error parameters}
    \label{TAB_GeometricalErrorParameters}
    \centering
    \TabGeometricalErrorParameters
\end{table}
\begin{figure}[tb]
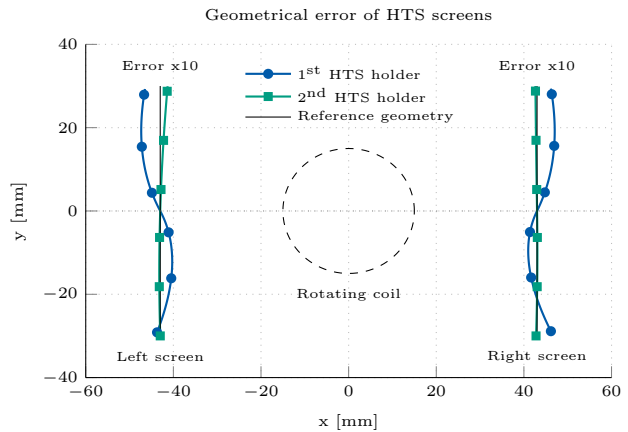

  \centering
    \def\fileName{Content/Data/Results_HALO_GeometricalError.csv}
	\FigLineHALOdeformation{\fileName}
	\caption{
	Graphical rendering of the geometrical error in the HTS screens, for both the version of the holder, as well as for an ideal HALO geometry, with perfectly parallel HTS screens. Errors are rendered with a factor ten amplification.}
	\label{FIG_LineHALOdeformation}
\end{figure}
\begin{figure*}[tb]
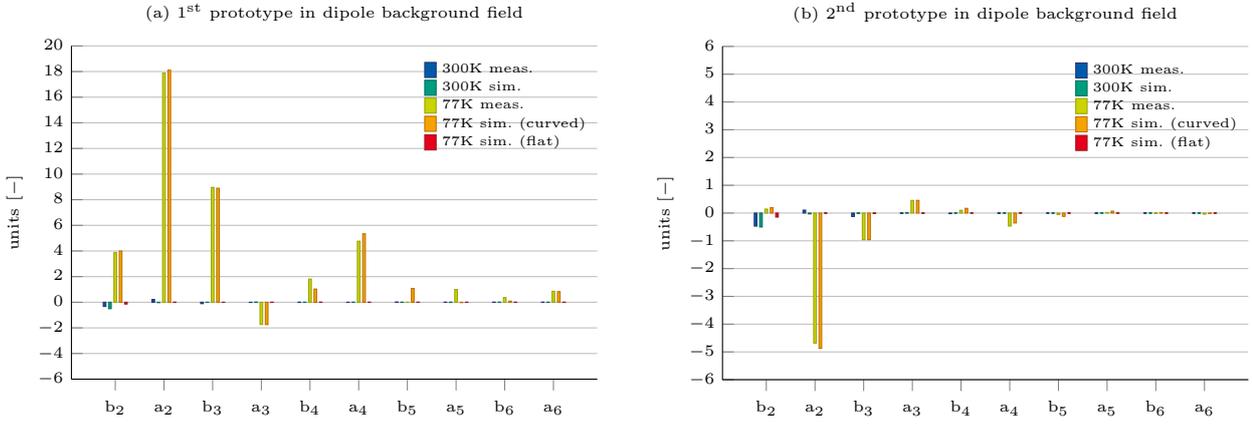

    \begin{minipage}[c]{0.5\textwidth}
            \centering
            \def\fileName{Content/Data/Results_HALO-V01.csv}	
            \def\figTitle{(a) $1^\RM{st}$ prototype in dipole background field}
            \def\ylimLo{-6}
            \def\ylimHi{20}
            \def\yStep{2} 
            \FigBarMultipolesHALO{\fileName}{\figTitle}{\ylimLo}{\ylimHi}{\yStep}
    \end{minipage}
    \hfill
    \begin{minipage}[c]{0.5\textwidth}
            \centering
            \def\fileName{Content/Data/Results_HALO-V02.csv}	
            \def\figTitle{(b) $2^\RM{nd}$ prototype in dipole background field}
            \def\ylimLo{-6}
            \def\ylimHi{6}
            \def\yStep{1} 
            \FigBarMultipolesHALO{\fileName}{\figTitle}{\ylimLo}{\ylimHi}{\yStep}
    \end{minipage}
    \caption{Measured and simulated magnetic field quality, given in units as a multipole expansion series, for the (a) first, and (b) second HTS holder prototype. Results are given for the experimental setup at both $\SI{300}{\kelvin}$ and $\SI{77}{\kelvin}$.}
	\label{FIG_BarMultipolesHALO}   
\end{figure*}

The field quality measurements of the HALO setup at $\SI{77}{\kelvin}$ without iron bars are used in the PSO algorithm (see Section~\ref{SUBSEC_GeometricalQuality}), obtaining a residual in the penalty function below 0.5 units. The geometrical error parameters calculated for the HTS screens are given in Table~\ref{TAB_GeometricalErrorParameters}, for both HTS holders. The first prototype suffers from relevant geometrical errors, up to a few millimeters. In the second prototype, the errors related to curvature and rotation are reduced by more than two orders of magnitude.

The geometrical-error parameters shape the two HTS screen prototypes as shown in Fig.~\ref{FIG_LineHALOdeformation}, where the geometrical error is graphically magnified by a factor ten. Note that from Fig~\ref{FIG_HALOVisualInspection}, the distortion shown in Fig.~\ref{FIG_LineHALOdeformation} is already apparent. The figure includes also the position of the rotating coil probe and the horizontal mid-plane. 

The first prototype is affected by substantial geometrical error, whereas the second prototype is much closer to the the reference geometry. A residual deformation of the left screen still persists, leading to a non-perfect parallelism between the tapes. 

\subsection{HTS Screens without Iron Bars}
    \label{SUBSEC_HTSScreensWithoutIronBars}
The magnetic field quality is evaluated with the experimental setup mounted in the magnet aperture within a dipole background field, without iron bars. 

\subsubsection{First Prototype at $\SI{300}{}$ and $\SI{77}{\kelvin}$}
%
Measurement and simulation results are given in Fig.~\ref{FIG_BarMultipolesHALO}a. Measurements at $\SI{300}{\kelvin}$ quantify the influence of magnetization and dynamic phenomena possibly occurring in the experimental setup within 0.5 units of field error. Measurements with the HTS screen at $\SI{77}{\kelvin}$ show an undesired self-field error dominated by the $\SCAl{a}{\RM{2}}$ and $\SCAl{b}{\RM{3}}$ components, with contributions from $\SCAl{b}{\RM{2}}$ and $\SCAl{a}{\RM{4}}$. When corrected for geometrical errors, both the measured and simulated THD factor for the first prototype is equal to 21. Note that when simulating without allowing for geometrical distortions, the simulated THD factor is almost zero, thus showing the importance of geometrical errors on the overall result.

\subsubsection{Second Prototype at $\SI{300}{}$ and $\SI{77}{\kelvin}$}
%
The same measurements and simulations are presented for the second prototype in Fig.~\ref{FIG_BarMultipolesHALO}b. The field error at room temperature is unchanged, whereas measurements at cold show five units of $\SCAl{a}{\RM{2}}$, with minor contributions below one unit, and a THD factor of five. Therefore, the overall self-field error is reduced by about a factor of four. Similar to the first prototype, the simulations with geometrical errors reproduce the measurement results quite well, which implies that the origin of the remaining field error is quite well understood, and further improvements in THD factor through improved manufacturing techniques seem quite feasible.

\subsubsection{Comparison}
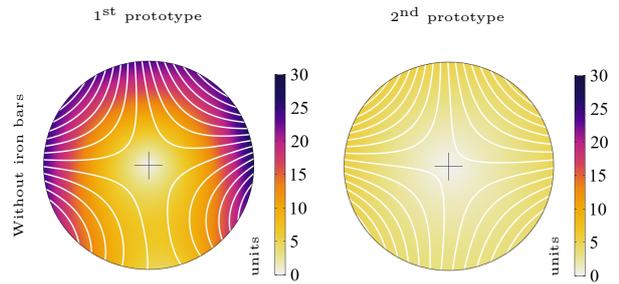
\begin{figure}[!h]
\raggedleft
    \begin{minipage}[c]{0.485\columnwidth}
    \centering
        \begin{tikzpicture}
            \node [inner sep=0pt] at (0,0) 
                {\includegraphics[width=4.0cm]{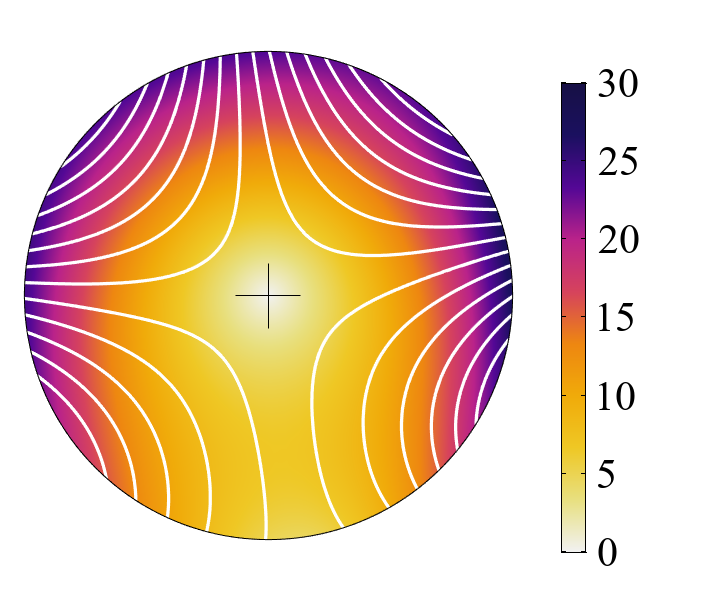}};
            \node [anchor=center,align=left,rotate=90] at (0.9, -1.2)
                {\scriptsize{units}}; 
            \node [anchor=center,align=center] at (-0.5, 2.0)
                {\scriptsize{$\RM{1^{st}}$ prototype}}; 
            \node [anchor=center,align=center,rotate=90] at (-2.2, 0)
                {\scriptsize{Without iron bars}}; 
        \end{tikzpicture}
    \end{minipage}
    \hfill
    \begin{minipage}[c]{0.485\columnwidth}
    \centering
        \begin{tikzpicture}
            \node [inner sep=0pt] at (0,0) 
                {\includegraphics[width=4.0cm]{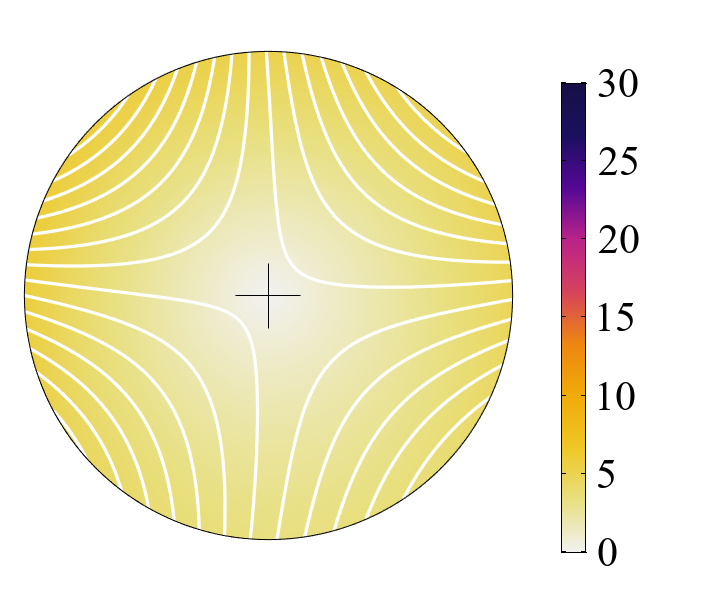}};
            \node [anchor=center,align=left,rotate=90] at (0.9, -1.2)
                {\scriptsize{units}}; 
            \node [anchor=center,align=center] at (-0.5, 2.0)
                {\scriptsize{$\RM{2^{nd}}$ prototype}}; 
        \end{tikzpicture}
    \end{minipage}
    \caption{Magnetic field error, in units, seen by the rotating coil probe and reconstructed from the measured multipole components. Results are shown for the two HALO prototypes at $\SI{77}{\kelvin}$, without iron bars.}
	\label{FIG_HALONoIronFieldQuality}    
\end{figure}

%
The net magnetic contribution, in units, provided by the HTS screens is presented in Fig.~\ref{FIG_HALONoIronFieldQuality} for the the rotating coil probe region. The field solution is reconstructed from the measured multipoles for both the version of the HTS holder, at $\SI{77}{\kelvin}$. The quadrupole field component is dominant in both cases, with a field gradient qualitatively higher with respect to the field on the right. 
\begin{figure}[tb]
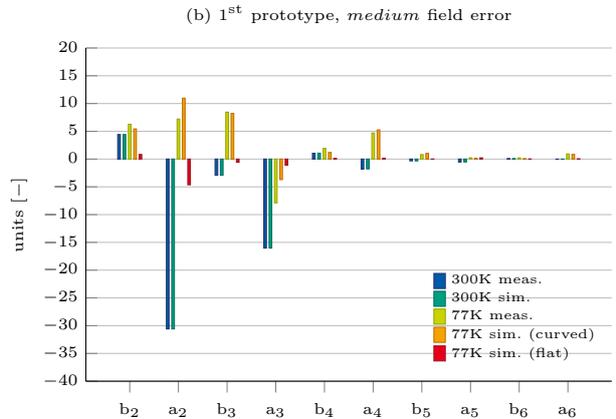

  \centering
    \def\fileName{Content/Data/Results_HALO-V01_mediumError.csv}
    \def\ylimLo{-40}
    \def\ylimHi{20}
        \def\figTitle{(b) $1^\RM{st}$ prototype, $medium$ field error}
    \def\IronPosLeft{+5}
    \def\IronPosRight{-20} 
    \FigBarMultipolesScenarios{\fileName}{\ylimLo}{\ylimHi}{\figTitle}{\IronPosLeft}{\IronPosRight}
	\caption{Measured and simulated magnetic field quality, given in units as a multipole expansion series. Results at $\SI{300}{\kelvin}$ are determined by the iron bars, whereas results at $\SI{77}{\kelvin}$ include also the HALO contribution.}
	\label{FIG_BarMultipolesIronHALOFirst}
\end{figure}

\subsection{HTS Screens with Iron Bars}
    \label{SUBSEC_HTSScreensWithIronBars}  
The iron bars are mounted in the magnet aperture, and the magnetic field quality is evaluated. Measurements at $\SI{300}{\kelvin}$ are influenced only by the iron bars, whereas at $\SI{77}{\kelvin}$ the HALO contribution is also included. 

\begin{figure*}[tb]
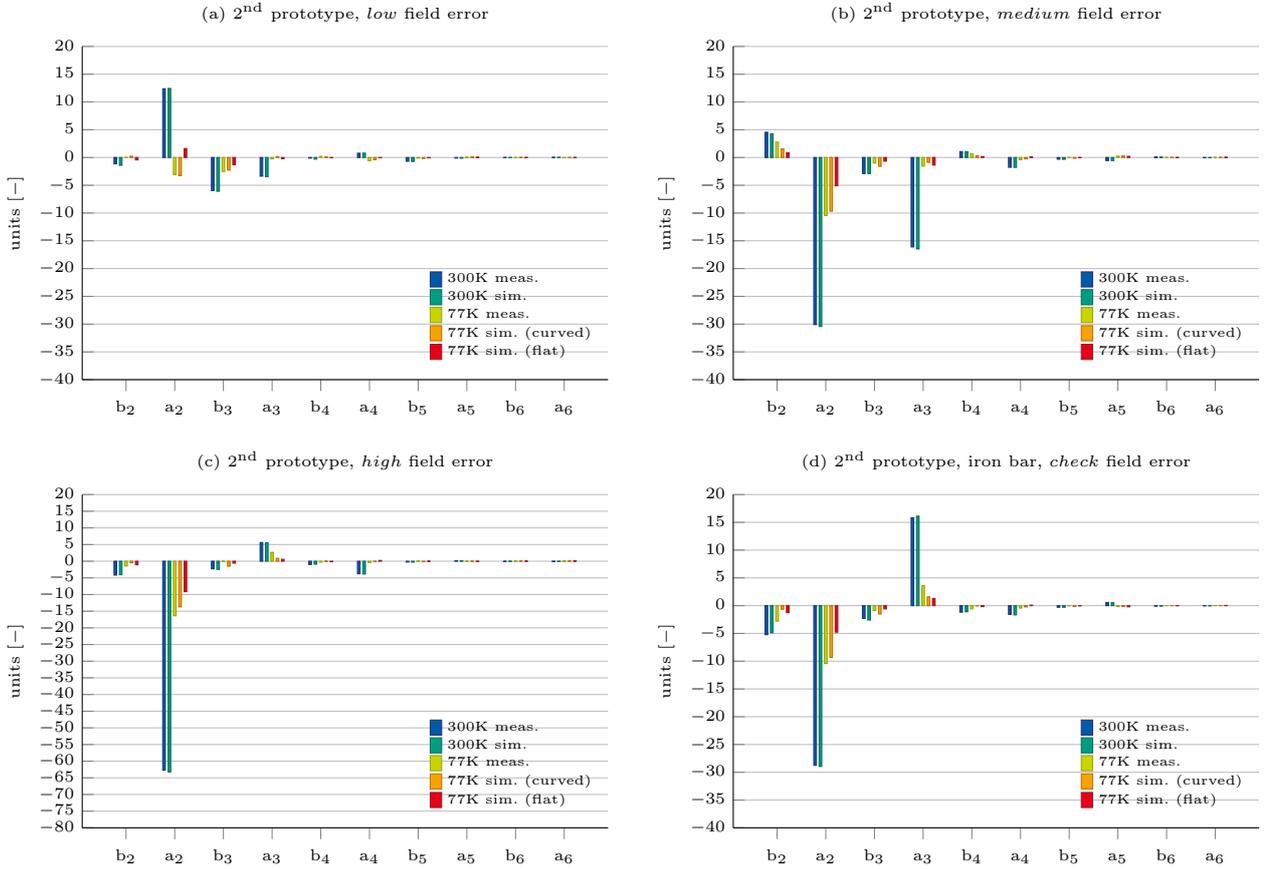

    \begin{minipage}[c]{0.5\textwidth}
        \centering
        \def\fileName{Content/Data/Results_HALO-V02_lowError.csv}
        \def\ylimLo{-40}
        \def\ylimHi{20}
        \def\figTitle{(a) $2^\RM{nd}$ prototype, $low$ field error}
        \def\IronPosLeft{+5}
        \def\IronPosRight{+0} 
        \FigBarMultipolesScenarios{\fileName}{\ylimLo}{\ylimHi}{\figTitle}{\IronPosLeft}{\IronPosRight}
    \end{minipage}
    \hfill
    \begin{minipage}[c]{0.5\textwidth}
        \centering
        \def\fileName{Content/Data/Results_HALO-V02_mediumError.csv} 
        \def\ylimLo{-40}
        \def\ylimHi{20}
        \def\figTitle{(b) $2^\RM{nd}$ prototype, $medium$ field error}
        \def\IronPosLeft{+5}
        \def\IronPosRight{-20}  
        \FigBarMultipolesScenarios{\fileName}{\ylimLo}{\ylimHi}{\figTitle}{\IronPosLeft}{\IronPosRight}
    \end{minipage}\\
    \begin{minipage}[c]{0.5\textwidth}
        \centering
        \def\fileName{Content/Data/Results_HALO-V02_highError.csv} 
        \def\ylimLo{-80}
        \def\ylimHi{20}
        \def\figTitle{(c) $2^\RM{nd}$ prototype, $high$ field error}
        \def\IronPosLeft{-20}
        \def\IronPosRight{-10} 
        \FigBarMultipolesScenarios{\fileName}{\ylimLo}{\ylimHi}{\figTitle}{\IronPosLeft}{\IronPosRight}
    \end{minipage}
    \hfill
    \begin{minipage}[c]{0.5\textwidth}
        \centering
        \def\fileName{Content/Data/Results_HALO-V02_checkError.csv} 
        \def\ylimLo{-40}
        \def\ylimHi{20}
        \def\figTitle{(d) $2^\RM{nd}$ prototype, iron bar, $check$ field error}
        \def\IronPosLeft{-20}
        \def\IronPosRight{+5}
        \FigBarMultipolesScenarios{\fileName}{\ylimLo}{\ylimHi}{\figTitle}{\IronPosLeft}{\IronPosRight}
    \end{minipage}
    \caption{    
    Measured and simulated magnetic field quality, given as a multipole expansion series. Results are shown, from top to bottom, for the (a) $low$, (b) $medium$ (c) $high$, and (d) $check$ error scenarios. Results at $\SI{300}{\kelvin}$ are determined by the iron bars, whereas results at $\SI{77}{\kelvin}$ include also the HALO contribution.
    }
	\label{FIG_BarMultipolesIronHALOSecond}
\end{figure*}
\begin{figure}[th!]
    \begin{minipage}[c]{0.485\columnwidth}
    \centering
        \begin{tikzpicture}[every text node part/.style={align=center}]
            \node [inner sep=0pt] at (0,0) 
                {\includegraphics[width=4.0cm]{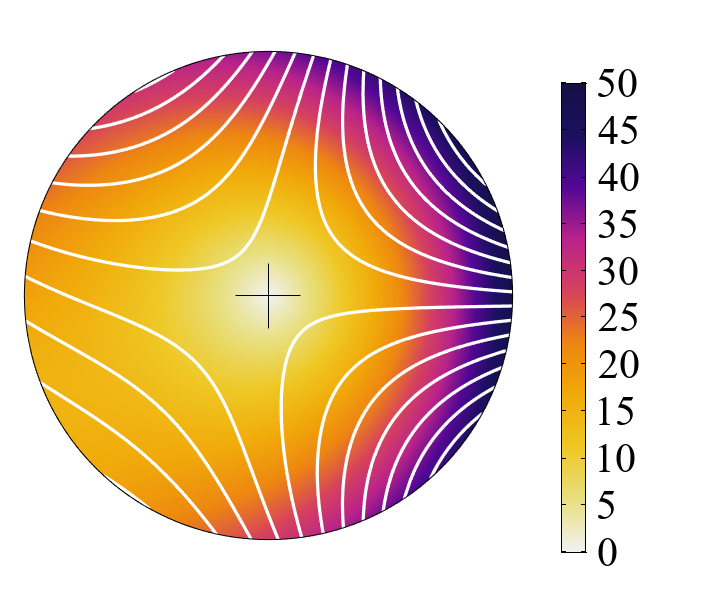}};
            \node [anchor=center,align=left,rotate=90] at (0.9, -1.2)
                {\scriptsize{units}}; 
            \node [anchor=center,align=center] at (-0.5, 2.0)
                {\scriptsize{$\SI{300}{\kelvin}$}}; 
            \node [anchor=center,align=center,rotate=90] at (-2.2, 0)
                {\scriptsize{$\RM{1^{st}}$ HALO, $medium$ error}}; 
        \end{tikzpicture}
        \vskip -0.1cm 
    \subcaption{}
    \end{minipage}
    \hfill
    \begin{minipage}[c]{0.485\columnwidth}
    \centering
        \begin{tikzpicture}
            \node [inner sep=0pt] at (0,0) 
                {\includegraphics[width=4.0cm]{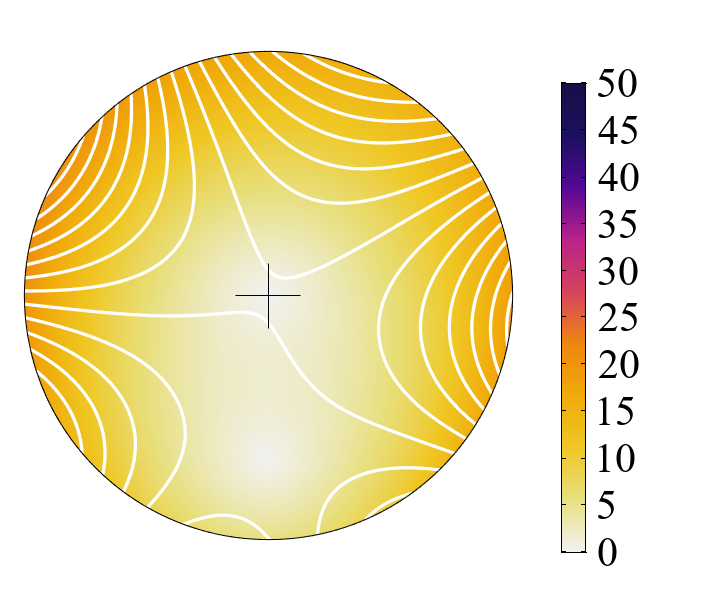}};
            \node [anchor=center,align=left,rotate=90] at (0.9, -1.2)
                {\scriptsize{units}}; 
            \node [anchor=center,align=center] at (-0.3, 2.0)
                {\scriptsize{$\SI{77}{\kelvin}$}};
        \end{tikzpicture}
        \vskip -0.1cm 
    \subcaption{\ \ \ \ \ }
    \end{minipage}
    \\
    \begin{minipage}[c]{0.485\columnwidth}
    \centering
        \begin{tikzpicture}
            \node [inner sep=0pt] at (0,0) 
                {\includegraphics[width=4.0cm]{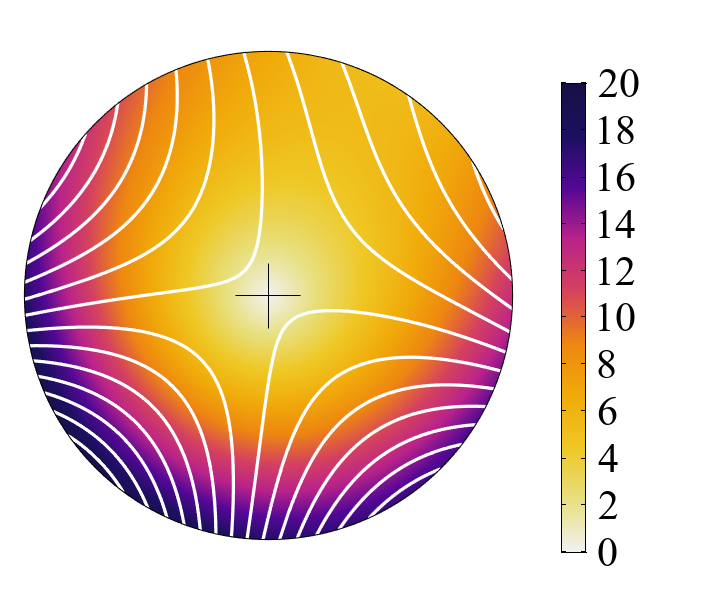}};
            \node [anchor=center,align=left,rotate=90] at (0.9, -1.2)
                {\scriptsize{units}}; 
            \node [anchor=center,align=center,rotate=90] at (-2.2, 0)
               {\scriptsize{$\RM{2^{nd}}$ HALO, $low$ error}}; 
        \end{tikzpicture}
        \vskip -0.1cm  
    \subcaption{}
    \end{minipage}
    \hfill
    \begin{minipage}[c]{0.485\columnwidth}
    \centering
        \begin{tikzpicture}
            \node [inner sep=0pt] at (0,0) 
                {\includegraphics[width=4.0cm]{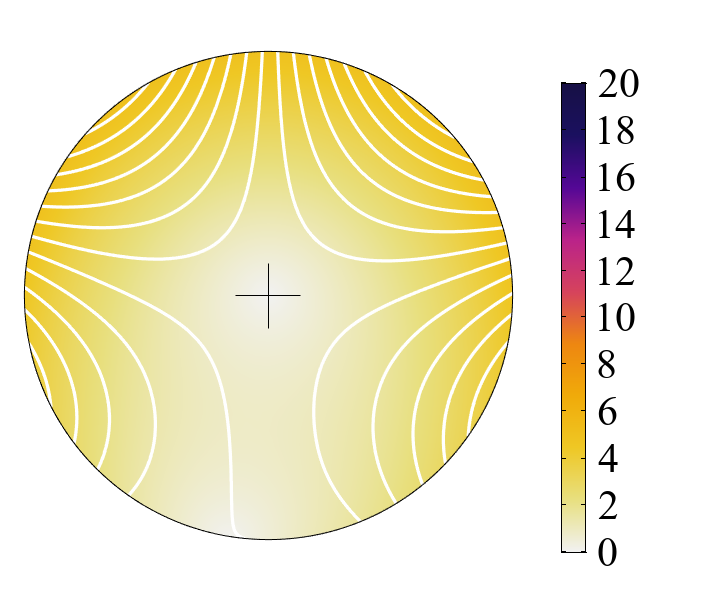}};
            \node [anchor=center,align=left,rotate=90] at (0.9, -1.2)
                {\scriptsize{units}}; 
        \end{tikzpicture}
        \vskip -0.1cm  
    \subcaption{\ \ \ \ \ }
    \end{minipage}
    \\
    \begin{minipage}[c]{0.485\columnwidth}
    \centering
        \begin{tikzpicture}
            \node [inner sep=0pt] at (0,0) 
                {\includegraphics[width=4.0cm]{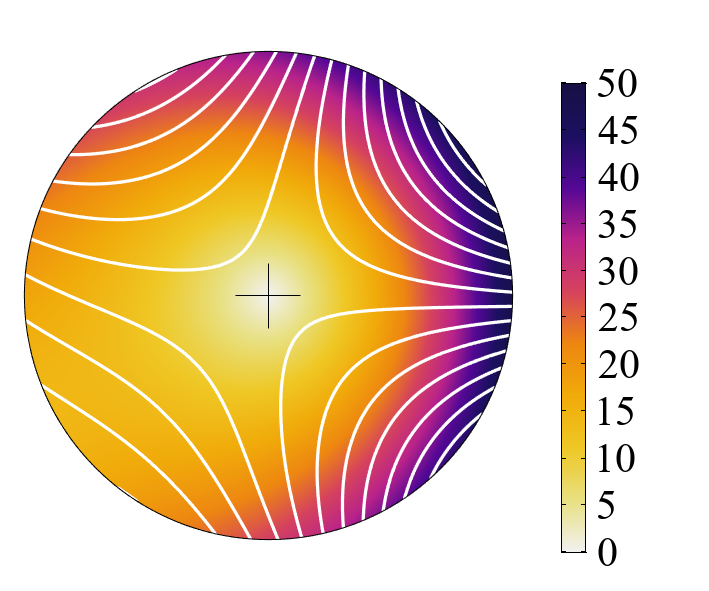}};
            \node [anchor=center,align=left,rotate=90] at (0.9, -1.2)
                {\scriptsize{units}}; 
            \node [anchor=center,align=center,rotate=90] at (-2.2, 0)
                {\scriptsize{$\RM{2^{nd}}$ HALO, $medium$ error}}; 
        \end{tikzpicture}
        \vskip -0.1cm  
    \subcaption{}
    \end{minipage}
    \hfill
    \begin{minipage}[c]{0.485\columnwidth}
    \centering
        \begin{tikzpicture}
            \node [inner sep=0pt] at (0,0) 
                {\includegraphics[width=4.0cm]{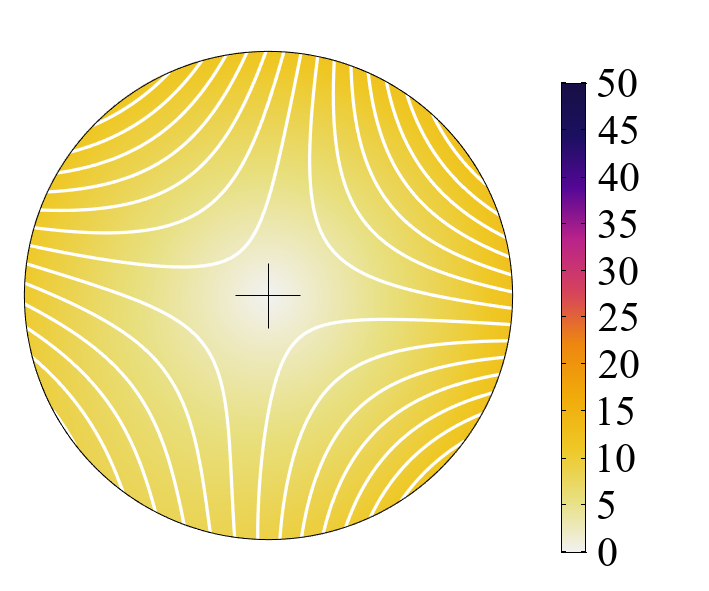}};
            \node [anchor=center,align=left,rotate=90] at (0.9, -1.2)
                {\scriptsize{units}}; 
        \end{tikzpicture}
        \vskip -0.1cm  
    \subcaption{\ \ \ \ \ }
    \end{minipage}
    \\
    \begin{minipage}[c]{0.485\columnwidth}
    \centering
        \begin{tikzpicture}
            \node [inner sep=0pt] at (0,0) 
                {\includegraphics[width=4.0cm]{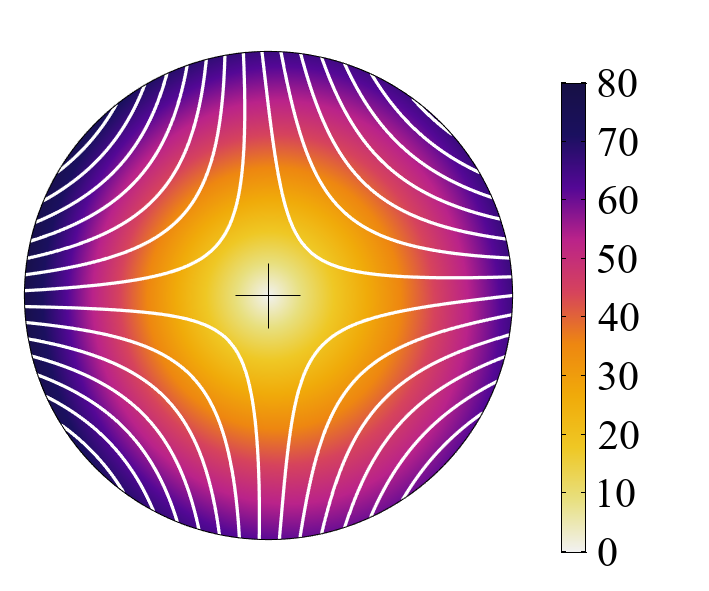}};
            \node [anchor=center,align=left,rotate=90] at (0.9, -1.2)
                {\scriptsize{units}}; 
            \node [anchor=center,align=center,rotate=90] at (-2.2, 0)
                {\scriptsize{$\RM{2^{nd}}$ HALO, $high$ error}}; 
        \end{tikzpicture}
        \vskip -0.1cm  
    \subcaption{}
    \end{minipage}
    \hfill
    \begin{minipage}[c]{0.485\columnwidth}
    \centering
        \begin{tikzpicture}
            \node [inner sep=0pt] at (0,0) 
                {\includegraphics[width=4.0cm]{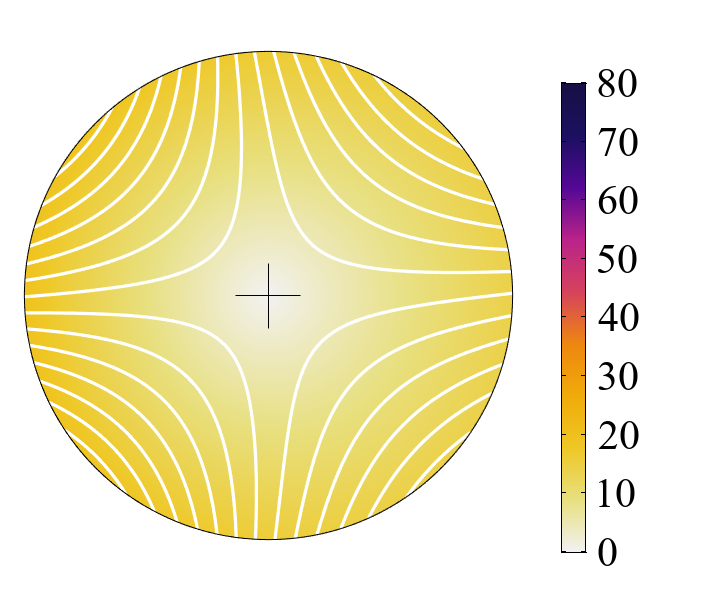}};
            \node [anchor=center,align=left,rotate=90] at (0.9, -1.2)
                {\scriptsize{units}}; 
        \end{tikzpicture}
        \vskip -0.1cm  
    \subcaption{\ \ \ \ \ }    
    \end{minipage}
    \\
    \begin{minipage}[c]{0.485\columnwidth}
    \centering
        \begin{tikzpicture}
            \node [inner sep=0pt] at (0,0) 
                {\includegraphics[width=4.0cm]{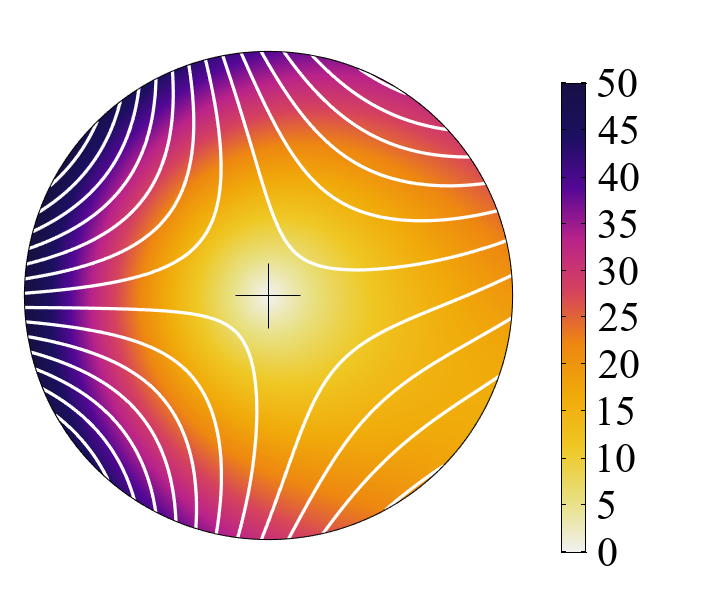}};
            \node [anchor=center,align=left,rotate=90] at (0.9, -1.2)
                {\scriptsize{units}}; 
            \node [anchor=center,align=center,rotate=90] at (-2.2, 0)
                {\scriptsize{$\RM{2^{nd}}$ HALO, $check$ error}}; 
        \end{tikzpicture}
        \vskip -0.1cm  
    \subcaption{}
    \end{minipage}
    \hfill
    \begin{minipage}[c]{0.485\columnwidth}
    \centering
        \begin{tikzpicture}
            \node [inner sep=0pt] at (0,0) 
                {\includegraphics[width=4.0cm]{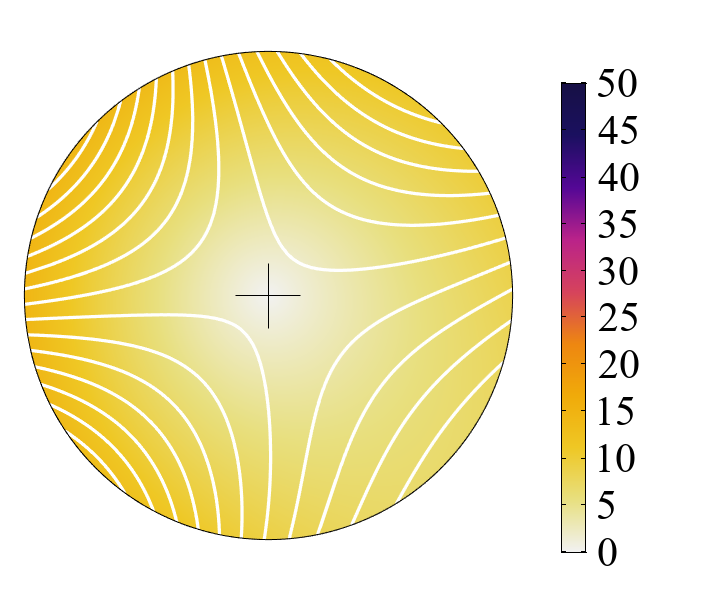}};
            \node [anchor=center,align=left,rotate=90] at (0.9, -1.2)
                {\scriptsize{units}}; 
        \end{tikzpicture}
        \vskip -0.1cm  
    \subcaption{\ \ \ \ \ }    
    \end{minipage}
    \caption{Magnetic field error, in units, as reconstructed from the measured multipole components. Results are shown at $\SI{300}{\kelvin}$ (left column) and $\SI{77}{\kelvin}$ (right column). Fig. (a,b) show the first HALO prototype in a $medium$ error scenario. Fig. (c,d), (e,f), (g,h), (i,j) show the second prototype in the $low$, $medium$, $hard$, and $check$ error scenarios, respectively.}
	\label{FIG_HALOIronFieldQuality}    
\end{figure}

\subsubsection{First Prototype at $\SI{300}{}$ and $\SI{77}{\kelvin}$}
%
The magnetic field quality is given in units as a multipole expansion series in Fig.~\ref{FIG_BarMultipolesIronHALOFirst}.  For the first prototype the iron bars are mounted accordingly to the $medium$ error scenario (see Table~\ref{TAB_FieldQualityTestCampaign}), creating 30 units of $\SCAl{a}{\RM{2}}$ and 15 units of $\SCAl{a}{\RM{3}}$, with minor contribution of $\SCAl{b}{\RM{2}}$. The HALO reduces the $\SCAl{a}{\RM{2}}$ $\SCAl{a}{\RM{3}}$ contributions, whereas the $\SCAl{b}{\RM{2}}$, $\SCAl{b}{\RM{3}}$ and $\SCAl{b}{\RM{4}}$ errors are increased. The increase for $\SCAl{b}{\RM{2}}$, $\SCAl{b}{\RM{3}}$ and $\SCAl{b}{\RM{4}}$ occurs due to the geometrical errors in the first HALO prototype (see section~\ref{SUBSEC_HTSScreensWithoutIronBars}).

\subsubsection{Second Prototype at $\SI{300}{}$ and $\SI{77}{\kelvin}$}
%
For the second prototype, the iron bars are mounted according to all the four scenarios described in Table~\ref{TAB_FieldQualityTestCampaign}. The field error due to the iron bars is characterized for the $low$ error scenario (Fig.~\ref{FIG_BarMultipolesIronHALOSecond}a) by 10 units of $\SCAl{a}{\RM{2}}$, with minor contributions of $\SCAl{b}{\RM{3}}$ and $\SCAl{a}{\RM{3}}$; for the the $medium$ error scenario (Fig.~\ref{FIG_BarMultipolesIronHALOSecond}b) by 30 units of $\SCAl{a}{\RM{2}}$ and 15 units of $\SCAl{a}{\RM{3}}$, with minor contribution of $\SCAl{b}{\RM{2}}$; for the the $high$ error scenario (Fig.~\ref{FIG_BarMultipolesIronHALOSecond}c) by 70 units of $\SCAl{a}{\RM{2}}$ and 10 units of $\SCAl{a}{\RM{3}}$ with minor contribution from $\SCAl{b}{\RM{2}}$ and $\SCAl{a}{\RM{4}}$. The $check$ error scenario (Fig.~\ref{FIG_BarMultipolesIronHALOSecond}d) introduces an error equal in magnitude to the $medium$ error scenario, but with inverted sign for the normal even-order and the skew odd-order multipoles. 

Once the HALO is superconducting, the field error is reduced. This observation holds true for for each scenario, and for every field multipole introduced by the iron bars. The reduction factor for the dominant multipoles is between three and four, higher for the components with higher magnitude. The high-order multipoles ($>4$) are left unperturbed, and are all within the noise floor of the background field. Simulation results show that curved and flat geometries are in qualitative agreement, with the screens working close to ideal conditions. Note that the geometrical error of the HALO stacks in the two prototypes is determined only once and in the absence of iron. With the subsequent addition of iron, and in spite of the absence of 're-fitting' for geometrical distortion, the measured and simulated results (Fig.~\ref{FIG_BarMultipolesIronHALOFirst} and ~\ref{FIG_BarMultipolesIronHALOSecond}) show a high degree of consistency, which is a strong indication of the excellent predictive value of the simulation model.

\subsubsection{Comparison}
%
The net magnetic contribution provided by the HTS screens is presented in Fig.~\ref{FIG_HALOIronFieldQuality} for the rotating coil probe region. Results are shown at $\SI{300}{\kelvin}$ (left column) and $\SI{77}{\kelvin}$ (right column). Figs. (a,b) show the first prototype in a $medium$ error scenario. Figs. (c,d), (e,f), (g,h), (i,j) show the second prototype in the $low$, $medium$, $hard$, and $check$ error scenarios. The color scales are consistent only by rows, as the field error in the four scenarios have different magnitude. It is possible to observe for all the scenarios, from the left to the right column, a reduction in the field error and an improvement in the homogeneity of the magnetic field distribution. 

\subsection{Field Error Cancellation}
    \label{SUBSEC_FieldErrorCancellation} 
%
\begin{figure}[tb]
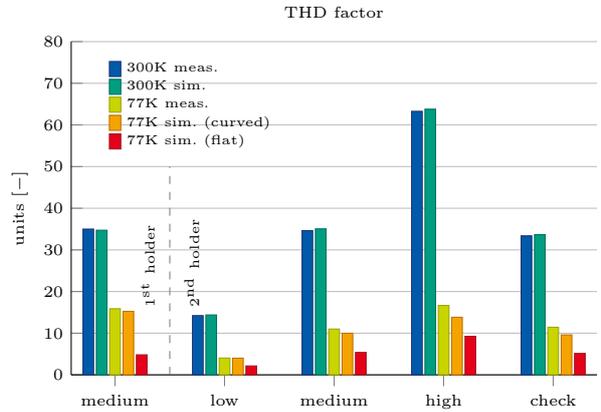

  \centering
    \def\fileName{Content/Data/Results_HALO-V01-02_THD.csv}
    \def\ylimLo{0}
    \def\ylimHi{80}
    \def\figTitle{THD factor}
	\FigBarTHD{\fileName}{\ylimLo}{\ylimHi}{\figTitle}
	\caption{
	Measured and simulated magnetic field quality, given in units as a THD factor. Results are shown for four different positions of the iron bars. The field error is shown for both the iron bars alone, and with the HALO. Simulation results are also shown for HTS screens with flat geometry.
	}
	\label{FIG_ResultsTHDAllScenarios}
\end{figure}

The THD factor, defined in~\eqref{EQ_THDfactor}, is calculated for the field error results for the first and second prototype (Fig.~\ref{FIG_BarMultipolesIronHALOFirst} and~\ref{FIG_BarMultipolesIronHALOSecond}). Results are reported in Fig.~\ref{FIG_ResultsTHDAllScenarios} for both measurements and simulations. The magnetic contribution of the HTS screens causes a reduction of the THD factor for each scenario, for both the first and the second prototype. The highest reduction is achieved for the scenario at high field error, where the contribution from the self-field error from the screen itself due to geometrical errors in the prototype geometries is less relevant.

The overall field-cancellation performance of the HTS screens is quantified by means of two efficiency parameters. The geometrical efficiency $\SCAl{\ITeta}{\RM{g}}$, together with the related geometrical quality factor $\RM{Q_{g}}$, measures the performance degradation caused by non ideal screen geometries. It is defined as
    \begin{align}
        \label{}
            \SCAl{\ITeta}{\RM{g}}(\VEC{\ITvarepsilon})=
            1-
            \frac{
            \SCAl{F}{\RM{d}}
            (\VECl{\RM{B}}{\RM{halo},\VEC{\ITvarepsilon}})
            }
            {
            \SCAl{F}{\RM{d}}
             (\VECl{\RM{B}}{\RM{iron}})
            }
            =
            \frac{\SCAl{Q}{\RM{g}}}{1+\SCAl{Q}{\RM{g}}},
    \end{align}
where $\VECl{\RM{B}}{\RM{halo},\VEC{\ITvarepsilon}}$ and $\VECl{\RM{B}}{\RM{iron}}$ represent the field in the dipole magnet, affected by the geometrical errors $\VEC{\ITvarepsilon}$ in the HTS screens, and by the iron bars. For high efficiency, the field error from the screens must be negligible with respect to the overall field error. For $\VEC{\ITvarepsilon}\to0$, thus for perfectly parallel and infinitely thin screens, $\SCAl{\ITeta}{\RM{g}}\to1$ and the screens reach the theoretical performance predicted by simulations with flat screens.

The magnetic efficiency $\SCAl{\ITeta}{\RM{m}}$, together with the related geometrical quality factor $\RM{Q_{m}}$, measures the overall field quality improvement after the field-error cancellation. It is defined as
    \begin{align}
        \label{EQ_defEfficiency}
            \SCAl{\ITeta}{\RM{m}}=
            1-
            \frac{
            \SCAl{F}{\RM{d}}
            (\VECl{\RM{B}}{\RM{both}})
            }
            {
            \SCAl{F}{\RM{d}}
             (\VECl{\RM{B}}{\RM{iron}})
            }
            =
            \frac{\RM{Q_{m}}}{1+\RM{Q_{m}}}.
    \end{align}
where $\VECl{\RM{B}}{\RM{both}}$ is the magnetic field in the dipole magnet with both HALO and the iron bars. The magnetic efficiency is influenced not only by the geometrical errors but also by the width, thickness, and position of the screens. A a complete cancellation of the magnetic field error corresponds to $\SCAl{\ITeta}{\RM{f}}=1$.

The performance parameters are reported in Table~\ref{TAB_HALOPerformancResults} for all the scenarios. The first prototype features $\ITeta_\RM{g}=0.4$ and $\RM{Q_{f}}=2.2$. A performance increase is observed for second prototype, achieving a $\ITeta_\RM{g}$ up to $90\%$ and $\SCAl{\ITeta}{\RM{f}}$ up to $75\%$, delivering a $\RM{Q_{f}}$ between 2.9 and 3.8. The field quality is improved by a factor of almost four, for the high field-error scenario. For the scenarios where $\RM{Q_{g}}<\RM{Q_{f}}$, the HALO self-field error provided a partial compensation of the field error due to the iron bars.

\begin{table}[tb]
    \caption{HALO performance results}
    \label{TAB_HALOPerformancResults}
    \centering
    \TabHALOPerformancResults
\end{table}
  

\section{Simulations}
    \label{SEC_Simulations}

The studied scenario is qualitatively represented by Fig.~\ref{FIG_HALO_PRINCIPLE}. The analysis is done for flat screens, considering an ideal geometry. The reference radius is retained from the experimental setup, and the screens are positioned $\SI{2}{\milli\meter}$ outside the magnet aperture. 

The screens are assumed to be operated at $\SI{4.5}{\kelvin}$, in a background dipole field of $\SI{10}{\tesla}$ affected by an error of 10 units of $\SCAl{b}{3}$. Due to the alignment of the tapes with the main field component, and the lower operational temperature with respect to the experiment, the lifting factor for the critical current increases. A constant value equal 10 is taken as a conservative assumption~\cite{senatore2015field}. 

\subsection{HTS Screens Geometry}
    \label{SUBSEC_HTSScreensGeometry}
The analysis starts from the screen geometry used in the prototype. Subsequently, the tapes per layer are increased from 5 to 8 (4 to 7 for the odd-order layers), up to $\SI{100}{\milli\meter}$-wide screens, and the number of layers is augmented from 4 to 16. It is worth noting that while the magnet coil design and the clearance of the magnet aperture may pose geometrical limits to the screen width, increasing the number of layers shall be eased by the negligible thickness of the tapes.

\begin{figure}[tb]
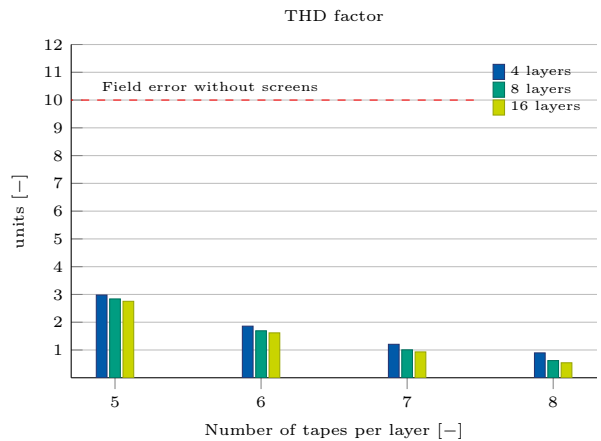

  \centering
    \def\fileName{Content/Data/Results_HALO-V02_Extrapolation_10T.csv}
	\FigBarResultsExtrapolationGeomTenTesla{\fileName}
	\caption{
	Simulated magnetic field quality, in units as a THD-index, as a function of the number of tapes per layer, and parametrized with the number of layers in the HTS screens.}
	\label{FIG_ResultsExtrapolationGeomTenTesla}
\end{figure}

The results, shown in Fig.~\ref{FIG_ResultsExtrapolationGeomTenTesla}, are given in terms of the THD factor as a function of the number of tapes per layer, and parametrized with the number of layers in the HTS screens. It is shown that the magnetic field error is reduced below one unit. If a field quality constraint is prescribed, the screen design can be scaled up to match the requirements. 
\begin{figure}[tb]
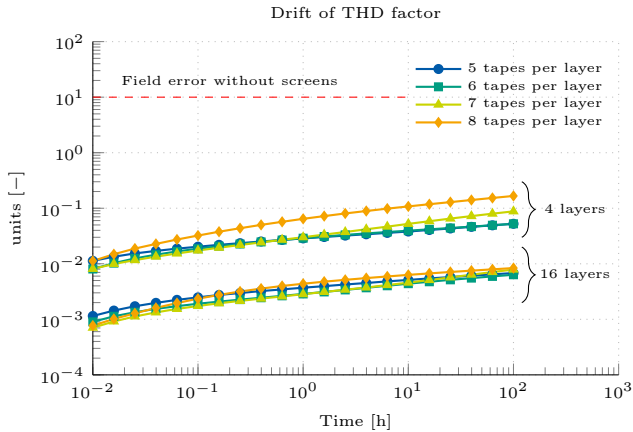

  \centering
    \def\fileName{Content/Data/Results_HALO-V02_Extrapolation_10T.csv}
	\FigLineResultsExtrapolationTimeTenTesla{\fileName}
	\caption{
	Simulated drift of the THD-index, in units, as a function of time. Results are shown for $\SI{100}{\hour}$ of stable operations, and are parametrized by both the number of tapes per layer, and the number of layers.
	}
	\label{FIG_ResultsExtrapolationTimeTenTesla}
\end{figure}

\subsection{Screening Currents Drift}
    \label{SUBSEC_ScreeningCurrentsDrift}
The HTS screens are simulated for up to $\SI{100}{\hour}$ of stable operations. The background field is increased from zero to $\SI{10}{\tesla}$, then the drift of the screening currents is simulated in steady-state field conditions. The drift is obtained as the increase of the THD factor with respect to the end of the field ramp. 

Results are shown in Fig.~\ref{FIG_ResultsExtrapolationTimeTenTesla} in terms of THD-factor drift as a function of time. The curves are parametrized by both the number of tapes per layer, and the number of layers. The increase of the THD factor is expected to remain below $\SI{0.01}{}$ units within $\SI{100}{\hour}$ for the 16-layers configuration. Results are conservative as the power law overestimates the field relaxation (see Sec.~\ref{SUBSEC_ConstitutiveEquation}. If a drift constraint is prescribed, the screen design can be scaled up to match the requirements.   
\section{Discussion} 
    \label{SEC_Discussion}
The measurements presented in Figs.~\ref{FIG_BarMultipolesIronHALOFirst} and~\ref{FIG_BarMultipolesIronHALOSecond} show in every scenario a relevant reduction of the field error, after the HALO has transitioned to the superconducting state. The cancellation effect is given by the magnetic contribution of screening currents induced in the HTS tapes. 

The cancellation occurs in all the multipole components, contributing to the homogenization of the magnetic field in the magnet aperture, as shown in Fig.~\ref{FIG_HALOIronFieldQuality}. The THD factor is reduced by a factor of three to four, as shown in Fig.~\ref{FIG_ResultsTHDAllScenarios}, depending on the field error scenario. The field-error cancellation provided by HALO brings a net improvement of the magnetic field quality. Therefore, the proof of concept can be considered successful.

The two holder prototypes reached about $70\%$ and $90\%$ of the field-error cancellation predicted by simulations assuming flat screens. The discrepancy is caused by geometrical deformations of the HTS screens due to mechanical tolerances which affected the manufacturing of the prototype. Such deformations have a detrimental influence on the field quality by introducing a self-field error which poses an upper limit to the HALO performance. The self-field error delivers a constant field contribution whose relevance decreases as the overall field error increases. 

In this work, the THD factor due to the self-field error was reduced to five units. It is expected that such error can be further reduced by suitable design choices, sufficiently tight mechanical tolerances and a precise manufacturing process. In detail, rigid HTS holders are recommended for the mechanical support of the tapes, e.g. rectangular blocks for dipole screens: the opposite faces can be accurately machined, and integrate the design of the screen with the one of the target application.

Numerical simulations for flat screens are found in qualitative agreement with measurements, although they overestimate the HALO performance. Geometrical deformations were identified in the first HTS holder prototype by means of a visual inspection, and were integrated in the model by means curved screens. The geometrical deformation magnitude was found by means of an optimization algorithm, leading to simulations in quantitative agreement with measurements. However, optimization results shall be used carefully, as they fit the model uncertainty, including the one about the superconducting properties of the tapes. Still, they are of help in understanding which geometrical errors affect the HTS screens performance.

Simulations at $\SI{4.5}{\kelvin}$ and in $\SI{10}{\tesla}$ dipole background field show that by increasing both the tape length and number of layers, the THD-index is reduced below one unit, leading to a nearly perfect cancellation of the field-error. The decay of the screening currents over time, and the consequent degradation of the field quality, is expected to remain within 0.1 units over 10 hours. 

The persistency of the screening currents is the cornerstone of the overall HALO technology. For this reason, the screening currents decay rate must cope with the field quality requirements in the target application. The field drift can be kept within specifications by choosing in the design phase appropriate features for HALO, such as the critical current of the tapes and number of layers in the screens.

The maximum field-error that can be canceled by screening currents is determined by the equivalent critical current of the HTS screens. This critical current can be increased either by increasing the layers of tapes, or decreasing the operational temperature. In order to integrate HALO in accelerator magnets, the combination of both the strategies is envisioned, as the magnetic fields are expected to be up to two orders of magnitude higher than in the experimental setup used for the proof of concept. For this reason, an operational temperature below $\SI{20}{\kelvin}$ is recommended.

\section{Conclusions and Outlook} 
    \label{SEC_Conclusions}

This paper presents the proof of concept for HALO (Harmonics-Absorbing Layered Object), a technology for field-error cancellation based on ReBCO tapes composing passive and self-regulating HTS screens. The working principle relies on the persistent magnetization from screening currents for shaping the magnetic field in a given region of space, e.g., the aperture of accelerator magnets. The method allows for a selective field error cancellation for both the dynamic and static contribution to the error.

The key-elements of the experimental setup are the reference dipole MCB24 from the magnetic measurement Laboratory at CERN for the reference field, two iron bars as sources of field error, the HTS screens providing the field error cancellation and a rotating coil probe for the field quality measurement. The field quality improvement is determined by differential measurements, without and with screening currents.

In the analysis of the results on the first prototype it was found that geometrical errors in this prototype had a detrimental effect on the measured field quality. The lessons learned from the first prototype were applied to the second prototype, and the observed geometrical distortion in the second prototype is found to be substantially better than in the first prototype. The field quality is measured in four field error configurations obtained with different positions for the iron bars. It was found that the HALO prototype provides a significant reduction of the THD factor associated to the field error, up to a factor four, reaching up to $90\%$ of the performance expected by numerical simulations. 

Measurements are compared with simulations. The analysis is carried out under magnetoquasistatic assumptions, using time-domain simulations based on a coupled $\VEC{A}$-$\VEC{H}$ formulation implemented in a 2D FEM model. Simulations provide the HALO theoretical performance in case of perfectly parallel HTS screens, and quantify the geometrical errors for the screens in the experimental setup, achieving quantitative agreement with measurements. 

Simulations shows that a complete error cancellation may be achieved by increasing the with and the thickness of the screen, for operational conditions compatible with accelerator magnets. At the same time, the field-quality drift due to the persistent currents decay can be kept within specifications by a suitable choice of design parameters. 

In an accelerator magnet, the HTS screens must be centered as close as possible around the beam vacuum chamber. The HALO technology is expected to provide the maximal benefit to magnets made of ReBCO tapes, as at low current the field quality is those magnets is expected to be heavily degrade by persistent magnetization phenomena. At the same time, HTS screens are applicable regardless of the technology used for the magnet, as long as they are kept in superconducting state. Moreover, the HALO technology might be a valid support also for all the applications beyond accelerators which need to satisfy stringent field quality requirements.



\section*{Acknowledgments} 
    \label{Acknowledgements}
  
This work has been sponsored by the Wolfgang Gentner Programme of the German Federal Ministry of Education and Research (grant no. 05E15CHA), and by Graduate School CE within the Centre for Computational Engineering at the Technische Universit{\"a}t Darmstadt. 

The authors would like to thank A.~Ballarino and S.~Hopkins for the procurement of the HTS tape, M.~Timmins for the support with the technical drawings, P.~Frichot for the procurement of the mechanical parts, M.~Liebsch, S.~Richter and T.~Nes for useful discussions about the design of the experimental setup, and L.~Fiscarelli for the support with measurements. The authors acknowledge the fruitful collaboration between CERN and the Technische Universit{\"a}t Darmstadt, within the framework of the STEAM collaboration project~\cite{STEAM2019website}.    
 

    {\footnotesize  
        \bibliography{main}
    }
 
              
\end{document}